\documentclass{aa}
\usepackage{graphicx}
\usepackage{txfonts}
\usepackage{natbib}
\usepackage{url}
\usepackage{longtable}
\usepackage{color}

\usepackage{lscape,rotating} %mettre en paysage 
\usepackage{color}

 \usepackage{threeparttable}

\begin{document}

   \title{Thermohaline instability and rotation-induced mixing} 
  \subtitle{ II-Yields of $^{3}$He for low- and intermediate-mass stars}
          
          \author{
          N. Lagarde\inst{1},
          C. Charbonnel\inst{1,2},
          T. Decressin\inst{1},\and
          J. Hagelberg\inst{1}
          }

   \institute{
              Geneva Observatory, University of Geneva, Chemin des Mailettes 51, 1290 Versoix, Switzerland\and
              IRAP, UMR 5277 CNRS and Universit\'e de Toulouse, 14, Av. E.Belin, 31400 Toulouse, France
             }
   \date{Submitted to A\&A on July 21, 2011 - Accepted on September 15, 2011 }

\authorrunning{N.Lagarde et al.  \titlerunning{Thermohaline instability and rotation-induced mixing. III-Yields of $^{3}$He}}

  \abstract
  % context heading (optional)
   {The $^{3}$He content of Galactic HII regions is very close to that of the Sun and the solar system, and only slightly higher than the primordial $^{3}$He abundance as predicted by the standard Big Bang nucleosynthesis. However, the classical theory of stellar evolution predicts a high production of $^{3}$He by low-mass stars, implying a strong increase of  $^{3}$He with time in the Galaxy. This is the well-known ``$^{3}$He problem".}
  % aims heading (mandatory)
   {We study the effects of thermohaline and rotation-induced mixings on the production and destruction of $^{3}$He over the lifetime of low- and intermediate-mass stars at various metallicities.}
  % methods heading (mandatory)
   {We compute stellar evolutionary models in the mass range 1 to 6~M$_{\odot}$ for four metallicities, taking into account thermohaline instability and rotation-induced mixing. For the thermohaline diffusivity we use the prescription based on the linear stability analysis, which reproduces Red Giant Branch (RGB) abundance patterns at all metallicities.
   %CC (Ulrich 1972, Charbonnel \& Zahn 2007b, Charbonnel \& Lagarde 2010). 
   Rotation-induced mixing is treated taking into account meridional circulation and shear turbulence.
   %CC  following the formalism of Zahn (1992) and Maeder \& Zahn (1998).
   We discuss the effects of these processes on internal and surface abundances of $^{3}$He and on the net yields.}
  % results heading (mandatory)
   {Over the whole mass and metallicity range investigated, rotation-induced mixing lowers the $^{3}$He production, as well as the upper mass limit at which stars destroy $^{3}$He. For low-mass stars, thermohaline mixing occuring beyond the RGB bump is the dominant process in strongly reducing the net $^{3}$He yield compared to standard computations. Yet these stars remain net $^{3}$He producers. }
  % conclusions heading (optional), leave it empty if necessary
   {Overall, the net $^{3}$He yields are strongly reduced compared to the standard framework predictions.}


   \maketitle

%===================================================   Introduction =============================================================================

\section{Introduction \label{intro}}

The classical theory of stellar evolution predicts a very simple Galactic 
destiny for $^3$He, dominated by the production of this isotope 
during the
Big Bang nucleosynthesis (BBN) and 
in stars  with initial masses lower than $\sim$ 3~M$_{\odot}$. 
In these objects, $^3$He is produced first through D-processing on the pre-main sequence 
and then through the pp-chain on the main sequence. This fresh $^3$He is then 
engulfed in the stellar convective envelope during the so-called first dredge-up when the stars move towards the red giant branch \citep{Iben67}. 
According to classical modelling\footnote{Classical (or standard) stellar models consider convection as the only mixing mechanism inside stars, and neglect all possible transport processes in stellar radiative regions.}, it survives the following stellar evolution phases before it is released in the interstellar matter through stellar wind and planetary nebula ejection \citep{Roodetal76,VaWo93,Dearbornetal96, Weissetal96, FoCh97}.
Two planetary nebulae, namely NGC~3242 and J320, whose estimated initial masses were slightly higher than that of the Sun, have been found to ``behave classically":
they are presently expelling freshly  synthesized elements among which is $^3$He with the amount predicted by classical stellar models \citep{Balser97,Balser99a,Balser06,Galli97}.

As a consequence, one expects a steep increase of $^3$He with time in the Galaxy 
with respect to its primordial abundance \citep[see e.g.][]{WilRoo94}, this latest quantity being well constrained through accurate determination of the baryon density of the Universe by recent cosmic microwave background experiments, most particularly from WMAP \citep{Spergel03,Dunkley09}, which has led to an unprecedented precision on the yields of standard BBN \citep{Coc04,Cyburt08}. Galactic HII regions in particular should be highly enriched in $^3$He because their matter content chronicles the result of billion years of chemical evolution since the Milky Way formation. 
Additionally, the present $^3$He/H abundance ratio is expected to be higher in the central regions of the Galaxy, where there has been more substantial stellar processing than in the solar neighbourhood.
However, the $^3$He abundance in HII regions that sample a large volume of the Galactic disk is found to be relatively homogeneous  \citep[][]{Roodetal79,Balser94,Balser99a,Bania97,Bania02,Bania10}; its average value is similar to that of the Sun at the epoch of its formation \citep[for references see][]{GeGl2010}, and only slightly higher than the WMAP+SBBN primordial abundance\footnote{The average abundance in Galactic HII regions is $^3$He/H=(1.9$\pm$0.6)$\times$10$^{-5}$ \citep{Bania02}. The protosolar and SBBN-WMAP values are $^3$He/H=(1.5$\pm$0.2)$\times$10$^{-5}$ \citep{GeGl98} and $^3$He/H=(1.04$\pm$0.04)$\times$10$^{-5}$ \citep{Coc04}.}.
No observational evidence is found therefore for the strong enrichment of this element in the Galactic history,  contrary to expectations from all chemical evolution models 
that take into account $^3$He yields from classical stellar models \citep[e.g.][]{Gallietal95, Oliveetal95,Tosi96}.

This is the well-known ``$^3$He problem" that could be solved if only $\sim$~10~$\%$ or less of all low-mass stars were actually releasing $^3$He, as predicted by the classical stellar theory \citep{Tosi1998,Tosi2000,Pallaetal00,Chiappinietal02,Romano03},
among which are NGC~3242 and J320.  
In other words, the lack of increase of the Galactic $^3$He abundance can be 
accounted for if most ($\geq$~90~$\%$) low-mass stars consume most of the $^3$He 
they produce during the main sequence before it can be released into the interstellar medium.
This calls for a physical process that is ignored by the classical theory of stellar evolution, but whose spectroscopic signatures have been revealed long ago at the surface of relatively bright low-mass red giants. 
In particular, one observes a sudden drop of the surface $^{12}$C/$^{13}$C ratio 
as low-mass stars evolve beyond the so-called luminosity bump on the red giant branch (RGB) well after the end of the first dredge-up
\citep{Gilroy89,GiBr91,Charbonnel94,ChBrWall98,Gratton00,Tautvaisiene00,Tautvaisiene05,Shetrone03,Pilachowski03,Spite06,ReLa07,CCNL10}.
This behaviour, which is not predicted by standard stellar modelling, 
appears to be almost universal. 
Indeed, about 96$\%$ of all low-mass bright red giant stars exhibit unexpectedly low $^{12}$C/$^{13}$C, regardless of whether they belong to the field, to open or globular clusters \citep{CharDoNa98}, or even to external galaxies \citep{Smith02,Geisler05}. 

This high number satisfies the Galactic requirements for the evolution of the $^3$He abundance 
because the mechanism responsible for the low values of $^{12}$C/$^{13}$C 
is also expected to lead to the depletion of $^3$He by a large factor in the stellar envelopes, as initially suggested by \citet[see also \citet{Charbonnel95,Hogan95,SaBo99,Eggleton06}]{Rood84}. This correlation was  recently confirmed by \cite{ChaZah07a} and \cite{Eggleton08}, 
who included the transport of chemicals caused by thermohaline mixing in stellar models \citep[see also][]{Stancliffe09,ChLa10}. In red giant stars, this double diffusive instability  (also called   "fingering convection"),  is induced by the mean molecular weight inversion created by the $^{3}$He($^{3}$He,2p)$^{4}$He reaction in the region between the hydrogen-burning shell and the convective envelope  \citep{ChaZah07a}\footnote{Attention to the local depression of $\mu$ occurring in RGB stars was drawn by \citet{Eggleton06}, although the peculiarity of this nuclear reaction that converts two particles into three was already pointed out by \citet{Ulrich71} in a different stellar context.}.
It sets in naturally as soon as stars reach the so-called luminosity bump on the RGB. However, and importantly in the $^3$He context, \cite{ChaZah07b} proposed that thermohaline mixing can be inhibited by a fossil magnetic field in red giant stars that are the descendants of Ap stars. As a consequence, these ``stubborn" objects are expected to enrich the ISM with $^3$He as predicted by standard models and as observed in very rare planetary nebulae. 
Their relative number is low \citep[of the order of 2-5$\%$ of all A-type stars, see references in][]{ChaZah07b}, which helps in principle to reconcile the long-standing problem of $^3$He overproduction on Galactic timescales with the measurements of $^3$He/H in planetary nebulae like NGC~3242 and J320. 

In order to validate the whole picture quantitatively, one needs to compute $^3$He yields  for stars of various masses and metallicities that contributed to the chemical evolution of the Galaxy, 
taking into account the various processes that may modify stellar nucleosynthesis.
This is the aim of the present work. 
In Paper I \citep{ChLa10} we presented evolution models for low- and intermediate-mass solar metallicity stars including the effects of both thermohaline and rotation-induced mixing. This study extended the former calculations by \cite{ChaZah07a} that focussed on low-metallicity stars and confirmed that thermohaline mixing is potentially the universal process that governs the photospheric composition of low-mass\footnote{We define low-mass stars as those that climb the red giant branch with a degenerate helium core; their initial mass is typically lower than 2-2.2~M$_{\odot}$ depending on metallicity. Intermediate-mass stars are those that ignite central helium-burning in a non-degenerate core at relatively low luminosity on the RGB, and finish their lives as C-O white dwarfs.} bright giant stars. 
In both papers we showed that when described with the prescription adopted by \citet[][ see \S~2.2]{ChaZah07a} this mechanism, whose efficiency on the RGB increases with decreasing initial stellar mass and metallicity, accounts very nicely for the observed behaviour of $^{12}$C/$^{13}$C, [N/C], and Li while efficiently lowering the $^3$He content in low-mass bright RGB stars of various metallicities. 
On the other hand, we also showed in Paper I that rotation-induced mixing on the main sequence 
changes the stellar structure so that it slightly re-enforces the effects of the thermohaline instability in low-mass stars and explains the observed features of CN-processed material in more massive evolved stars that do not undergo thermohaline mixing on the RGB. 
Last but not least, in Paper I thermohaline mixing was found to lead to additional $^3$He depletion associated to Li production in all the solar-metallicity models that we computed along the thermal pulse phase on the asymptotic giant branch (TP-AGB), confirming the previous findings by  \cite{Stancliffe10} for low-metallicity low-mass stars. This accounts beautifully for the Li behaviour in Galactic oxygen-rich AGB variables.

Based on these successes we presently extend the computations at different metallicities within the same framework 
to quantify the impact of thermohaline and rotation-induced mixings on the yields of $^{3}$He in low- and intermediate-mass stars that are classicaly considered to be net producers of $^{3}$He.
In \S~2 we briefly recall the assumptions and input physics of our stellar models. In \S~3 we present the theoretical predictions for $^3$He nucleosynthesis both in standard (or classical) models and in models where thermohaline and rotation-induced instabilities are taken into account. The corresponding yields of $^{3}$He are given and discussed in \S~4.
The impact of our new $^{3}$He yields on Galactic chemical evolution will be presented in a separate paper.

%===================================================   fin Introduction =============================================================================

\section{Input physics of the stellar models
 \label{hypothesis}}
 
 \subsection{Basic assumptions}

We present predictions for stellar models computed with the code STAREVOL (V3.00) for a range of initial masses between 0.85 and 6~M$_{\odot}$\footnote{We do not compute models for more massive stars (i.e., with initial mass higher than $\sim$ 6~M$_{\odot}$) that are anyway considered as net $^3$He destroyers within the standard framework since a large amount of their material is processed at temperatures high enough to burn any present $^3$He to $^4$He or beyond \citep{Dearbornetal86}.} and for four metallicities Z=0.0001, 0.002, 0.004, and 0.14, which correspond to [Fe/H]= $-2.16$, $-0.86$,  $-0.56$, and 0\footnote{The ratio measuring the enrichment in $^{4}$He reported  to the enrichment in heavy elements in the Galaxy until the birth of the Sun is taken equal to $\Delta Y / \Delta Z =1.29$ (see Lagarde et al., Paper III, in prep.). We assume $[\alpha/Fe]=0$ at all metallicities, which has a negligible impact on the final yields of $^{3}$He. Indeed, for the [1.5~M$_{\odot}$; Z=0.0001] models  the $^3$He yields are only $\sim$4.6-4.7\% (standard and thermohaline cases respectively) higher when [$\alpha$/Fe] is increased to 0.3.}. 
For each given stellar mass and metallicity, models are computed from the beginning of the pre-main sequence (along the Hayashi track) up to the end of the second dredge-up on the early-AGB with different assumptions: (1) standard (no mixing mechanism other than convection), (2) with thermohaline mixing only, and (3) with both thermohaline and rotation-induced mixing. Selected models are pursued along the TP-AGB up to the end of the superwind phase  (see \S \ref{casesABC} and \ref{nucleosurface} for details). The general evolution and detailed characteristics of the ensemble of our models are presented in Paper III (Lagarde et al., in prep.) of this series.

We refer to \citet{ChLa10} and to Paper III for a detailed description of the physical ingredients of the models. 
For the primordial D/H at all Z we assumed the WMAP-SBBN value of $2.6.10^{-5}$ \citep{Coc04}, which is higher than the protosolar value 2.1$\pm$0.5$\times$10$^{-5}$ \citep{GeGl98}.In addition, the initial value of $^3$He/H is assumed to vary with the metallicity from $^{3}$He/H=1.17$\times$10$^{-5}$ for Z=0.0001 to $^{3}$He/H=1.31$\times$10$^{-5}$ for Z=0.014.

For the nuclear reactions involving $^3$He we used the nominal rates from NACRE compilation \citep{Angulo99} ; the corresponding uncertainties are less than 6$\%$ (except for D(H,$\gamma)^{3}$He, for which it amounts to 40$\%$) and are therefore not affecting our conclusions.
Convection was treated within the standard mixing length theory with an $\alpha$-parameter taken equal to 1.6, and no overshooting or semi-convection was included. We assumed instantaneous convective mixing, except when hot-bottom burning occurs, which requires a time-dependent convective diffusion algorithm as developed in \cite{FoCh97}. The treatment of transport processes in radiative regions is described in \S\ref{transport}.
For mass loss we used the \citet{Reimers75} formula (with $\eta_{R}$=0.5) up to central helium exhaustion and then shifted to the \citet{VaWo93} prescription on the AGB.

\subsection{Transport processes in radiative regions \label{transport}}

\begin{figure*} 
	\centering
	~\hfill Standard models\hfill\hfill Rotating models\hfill ~ \\
		\includegraphics[angle=0,width=9cm]{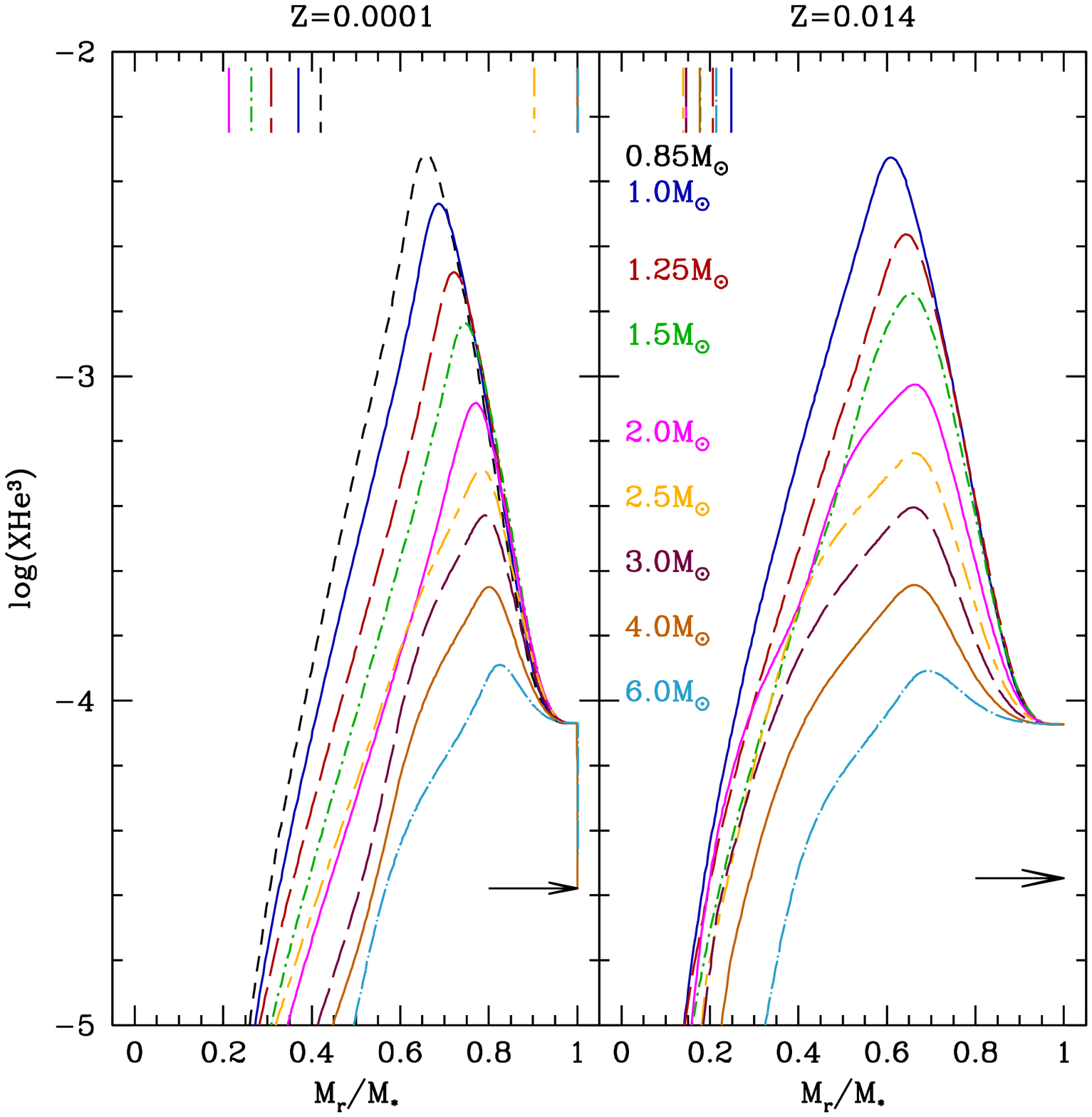}
		\includegraphics[angle=0,width=9cm]{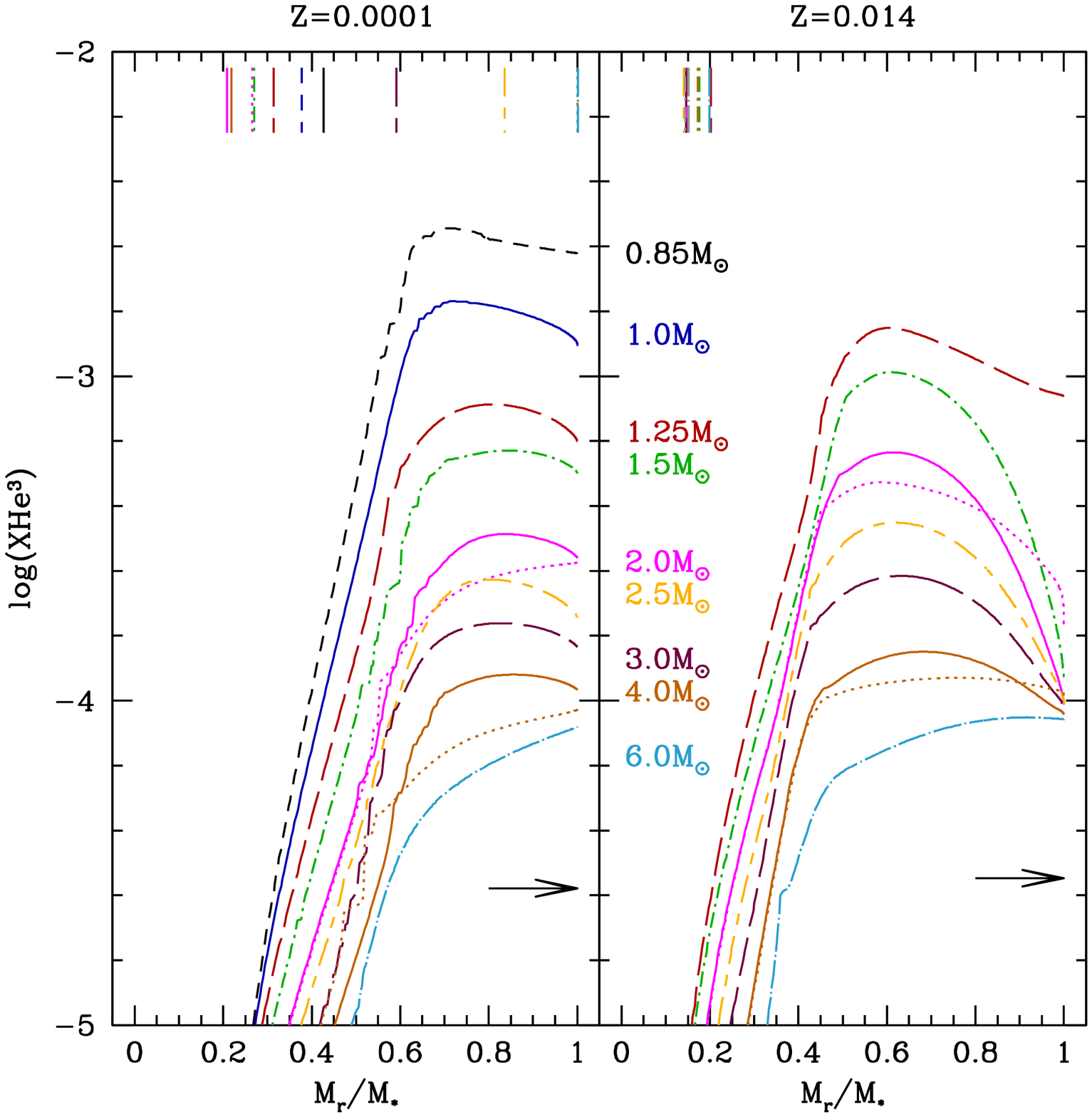}
	  \caption{$^3$He profile (in mass fraction) at the main-sequence turnoff for models of various masses as indicated and for two values of the metallicity (Z=$10^{-4}$ and Z$_{\odot}$ in the left and right subpanels respectively). The horizontal arrows indicate the initial $^3$He content assumed at stellar birth. The vertical lines show, in each case, the maximum depth reached by the convective envelope during the first dredge-up. {\sl(Left)} Standard models. { \sl (Right)} Models including rotation-induced mixing, assuming an initial rotation velocity equal to 45$\%$ of the critical velocity on the zero age main sequence; for the [2, 4~M${\odot}$; Z${\odot}$ and Z=$10^{-04}$] rotating models, the dotted curves correspond to predictions assuming V$_{\rm ZAMS}$/V$_{\rm crit}$$\sim$0.90}
	\label{fig:3HepeakvsMvsZ}
\end{figure*}
 
\begin{figure} 
	\centering
		\includegraphics[angle=0,width=9cm]{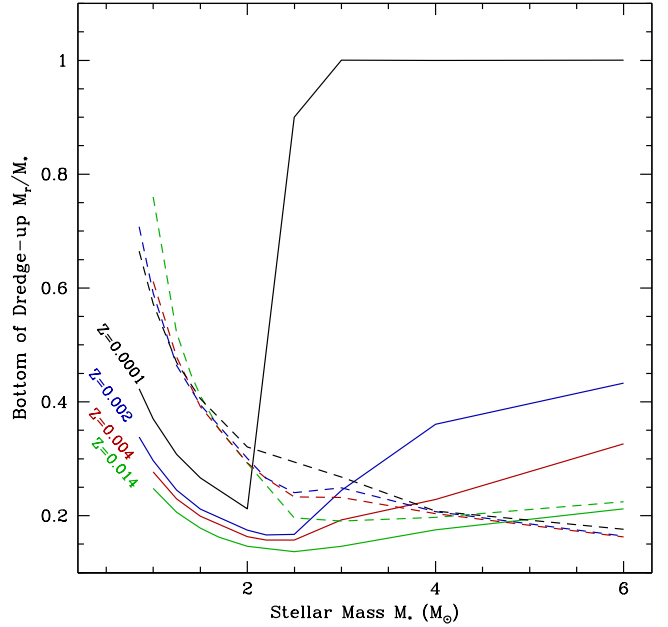}
	  \caption{Maximum depth in mass of the convective envelope relative to the total stellar mass reached during first and second dredge-up (solid and dashed lines respectively) as a function of initial stellar mass and for the different metallicities, as indicated by the colours of the curves. }
	\label{fig:Bottom_CE}
\end{figure}

\begin{figure*} 
	\centering
	~\hfill \hfill Standard models\hfill\hfill\hfill Thermohaline +Rotating models\hfill ~ \\
		\includegraphics[angle=0,width=9cm]{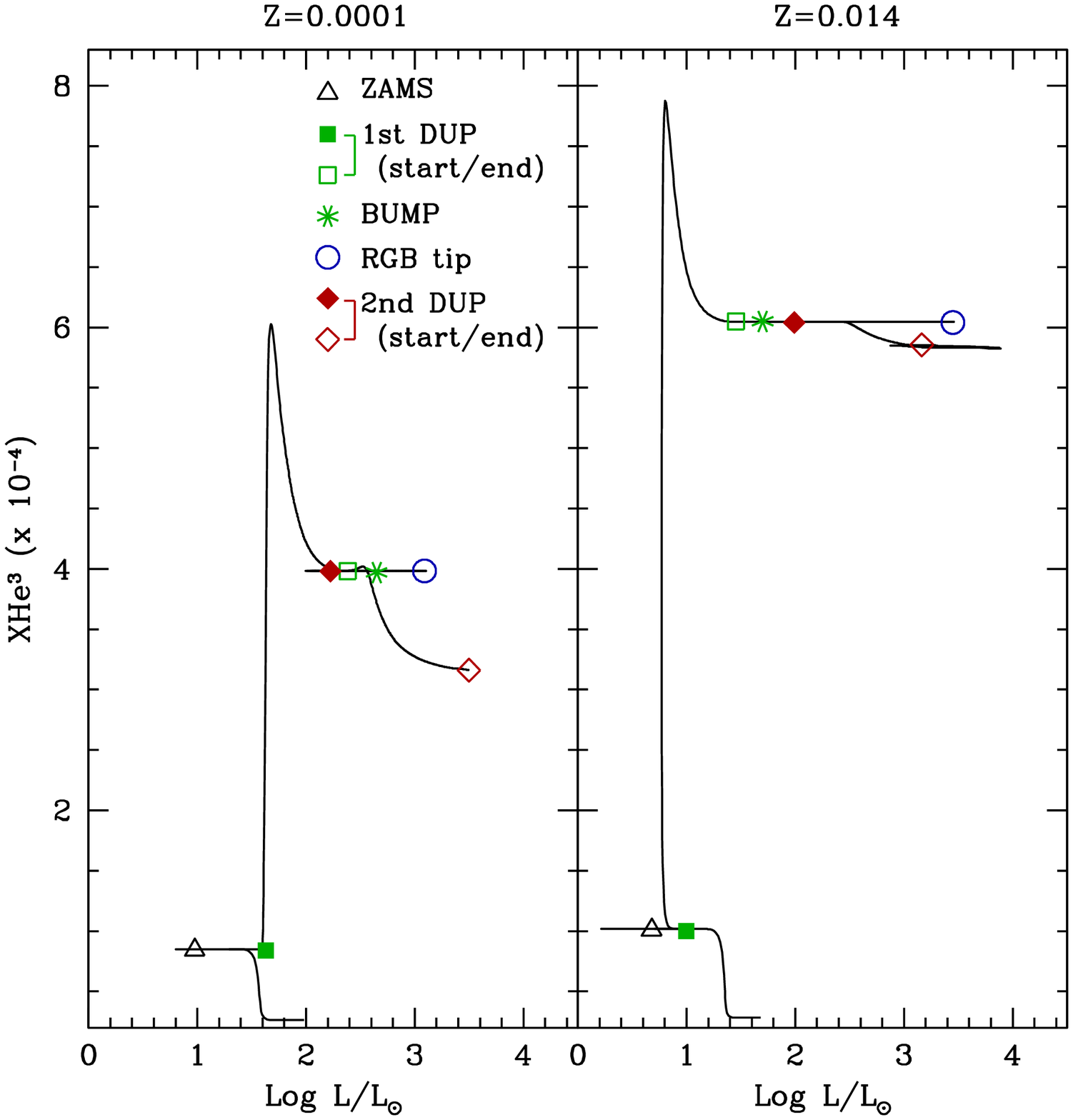}
		\includegraphics[angle=0,width=9cm]{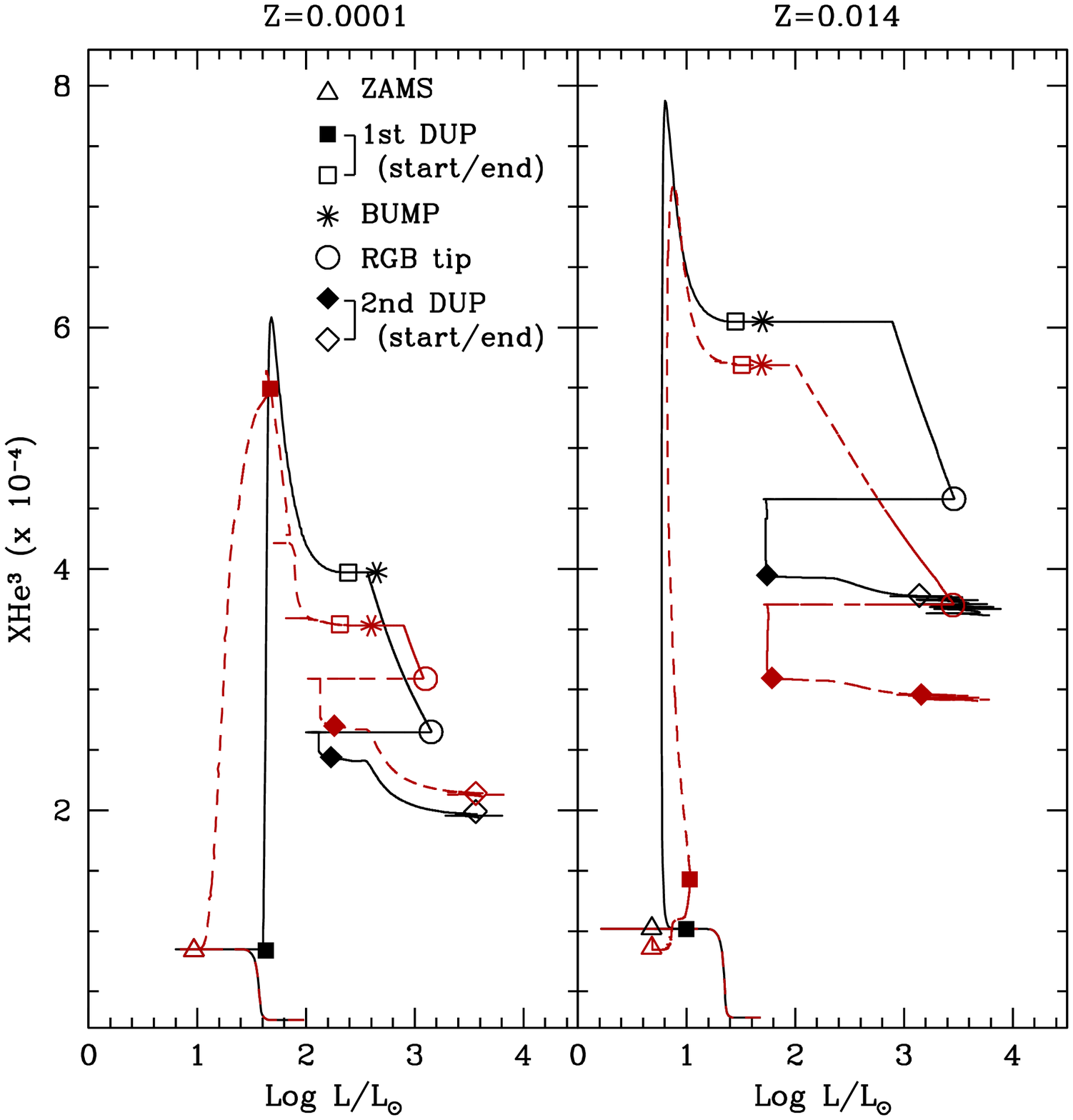}
	  \caption{Evolution of the surface abundance of $^{3}$He as a function of stellar luminosity for 1.5~M$_{\odot}$ models with Z=$10^{-4}$ and Z$_{\odot}$ as indicated. Main evolution steps are pointed out with different symbols. {\sl (Left)} Standard case computed up to the end of the second dredge-up only. {\sl (Right)} Models including either thermohaline mixing only (black solid lines), or both thermohaline and rotation-induced mixing (red dashed lines; $V_{ZAMS}/V_{crit}=0.45$); at  Z$_{\odot}$ the non-standard models are both computed up to the AGB tip.}
	\label{fig:He3_1p5_0001_014}
\end{figure*}

\subsubsection{Thermohaline instability}
For the thermohaline diffusivity we used the prescription advocated by  \cite{ChaZah07a} that beautifully reproduces RGB abundance patterns at all metallicities (see \S1). 
It is based on the linear stability analysis by \cite{Ulrich72} of the Boussinesq equations that describe motion in a nearly incompressible stratified viscous fluid, and includes the \cite{Kippen80} terms for a non-perfect gas \citep[for more details see][]{ChaZah07a,ChLa10}. 
The corresponding favoured geometry of the instability cells  is that of long thin fingers with the aspect ratio 
(i.e., maximum length relative to their diameter, $\alpha= l/d=5$ to 6) first obtained by \cite{Ulrich72} and supported by the \cite{Krish03} laboratory experiments. 
Although it is quite successful in reproducing the abundance data for evolved stars over a wide range in both mass and metallicity as shown in our previous studies \citep[see also][]{Denissenkov10}, this value for $\alpha$ turns out to be $\sim$ 5-10 times higher than 
that obtained by current 2D and 3D numerical simulations of thermohaline convection \citep{Denissenkov10, DenissenkovMerryfield10,RosenblumGaraudetal11,Traxleretal11}.  
However, these simulations are still far from the stellar regime. 
Indeed, even in the ``best" case they are run at moderatly low values of the Prandtl number  \citep[1/3 to 1/30][]{Traxleretal11}, which is several orders of magnitude away from stellar conditions. In the outer radiative wing of the hydrogen-burning shell of a low-mass RGB star, the Prandtl number varies indeed from $\sim$ $3 \times 10^{-6}$ to $3 \times 10^{-7}$. The same difficulty arises when the density ratio assumed in the simulations is concerned (up to 7 maximum, compared to $\sim 2  \times 10^3$ in the RGB case).
This casts some doubt on the accuracy and applicability in the stellar regime of the corresponding empirically determined transport laws. Additionally (and not surprisingly), the use in stellar models of  the corresponding low $\alpha$ values  precludes surface abundance  variations on the RGB   as shown by \citet{Wachlin11}. In our view, this urgently calls for a numerical exploration of low P\'eclet values.  
Before this becomes available, 
we keep to the analytical prescription that is able to describe the observational data on the RGB 
at all metallicities so well, and thus perform our computations with $\alpha = 6$ as in \cite{ChaZah07a} and \cite{ChLa10}.

\subsubsection{Rotation-induced mixing}
Rotation-induced mixing is treated as in Charbonnel \& Lagarde (2010; see also \citealt{Decressin09}) using the complete formalism developed by \cite{Zahn92} and \cite{MaeZah98} that takes into account the evolution of angular momentum and chemicals under the combined action of meridional circulation and shear turbulence. This complete treatment is applied up to the RGB tip or up to the second dredge-up for stars with masses below or above 2~M$_{\odot}$, respectively.

For all rotating models the initial rotation velocity on the zero age main sequence, V$_{\rm ZAMS}$, is chosen equal to 45$\%$ of the critical rotation velocity of the corresponding model at that evolution point, V$_{\rm crit}$. This corresponds to the mean observed values for these stars (see more details in Paper III and Ekstr\"om et al., submitted).
A couple of models were computed with higher initial rotation velocities to quantify the impact on the $^3$He yields (see the values in Table 1 and 4 and \S3 and 4 for a discussion).

\subsection{Computation of the $^3$He yields \label{casesABC}}

Except in a few cases that will be discussed in detail in \S~\ref{nucleosurface}, the standard computations were stopped at the end of the second dredge-up on the early-AGB (label A in the last column of Tables~1 to 4), and the net $^3$He yields are extrapolated as described below.  
For some of the non-standard models, however, we pursued the computations on the TP-AGB including the effects of thermohaline instability. For rotating stars and for one classical star (6.0$M_{\odot}$) that currently undergoes a hot-bottom burning process (HBB), models were computed until the complete consumption of  $^3$He in the stellar envelope if it is reached on the TP-AGB phase (label B in Tables~1 to 4 ). For non-standard models that do not undergo HBB, we stopped the computations either at the end of the second dredge-up on early AGB or at the end of the superwind phase at the AGB tip (labels A and C in Tables~1 to 4). 

In cases A and B the net $^3$He yields have to be extrapolated from the $^3$He content of the convective envelope in the last model computed along the corresponding evolutionary sequence. 
To estimate the mass lost from that point up to the AGB tip, we used the relation by \citet{Dobbieetal06} between the initial stellar mass and the mass of the white dwarfs (M$_{\rm final}=0.289{\rm M_{initial}}+0.133$).
In addition, we assume that the $^{3}$He abundance does not change at the stellar surface and consequently in the stellar wind during the final evolution on the TP-AGB. This assumption is reasonable for stars with initial masses lower than, or equal to  $\sim$ 3~M$_{\odot}$ that do not undergo hot-bottom burning on the TP-AGB, and in which the thermohaline instability leads only to very modest $^3$He depletion during that phase, as will be discussed below. The impact of these hypotheses will be quantified in \S~\ref{nucleosurface} and \ref{yields}.

\section{$^3$He nucleosynthesis and surface abundance \label{nucleosurface}}

\subsection {Standard predictions for the production and destruction of $^3$He in low- and intermediate-mass stars\label{standardmodels}}

\subsubsection{STAREVOL standard predictions \label{starevolstandard}}

While on the pre-main sequence, low- and intermediate-mass stars are converting pristine D into $^3$He via proton-capture at relatively low temperatures ($\geq$ 6 $\times 10^5$K ) within their contracting interior. 
Then on the main sequence a peak of fresh $^3$He builds up in  these objects as a result of the competition within the pp-chain between the production reactions (namely p(p, D) followed by D(p, $^3$He)) and the destruction ones (i.e., $^3$He($^3$He, $^4$He), $^3$He($^4$He, $7$Be)). The position and size of the peak depends on the stellar mass and metallicity, as can be seen in Fig.\ref{fig:3HepeakvsMvsZ} (left panels for the standard models). Lower initial stellar mass at a given metallicity as well as higher metallicity at a given stellar mass both result in a higher $^3$He abundance outside the regions of complete pp-processing caused by a longer main-sequence lifetime and by the dominance of the pp-chains with respect to the CNO-cycle. 
Additionally, lower initial mass and higher metallicity imply a more extended and deeper (in mass) $^3$He-rich region owing to lower temperatures at given depths as well as flatter d$T$/d$M_r$ gradients within the stellar interior during the evolution on the main sequence.
Consequently, the net production of $^3$He during that phase increases with decreasing stellar mass and increasing metallicity. 

\begin{figure*} 
	\centering
	~\hfill \hfill Standard models\hfill\hfill\hfill Thermohaline +Rotating models\hfill ~ \\		
		\includegraphics[angle=0,width=9cm]{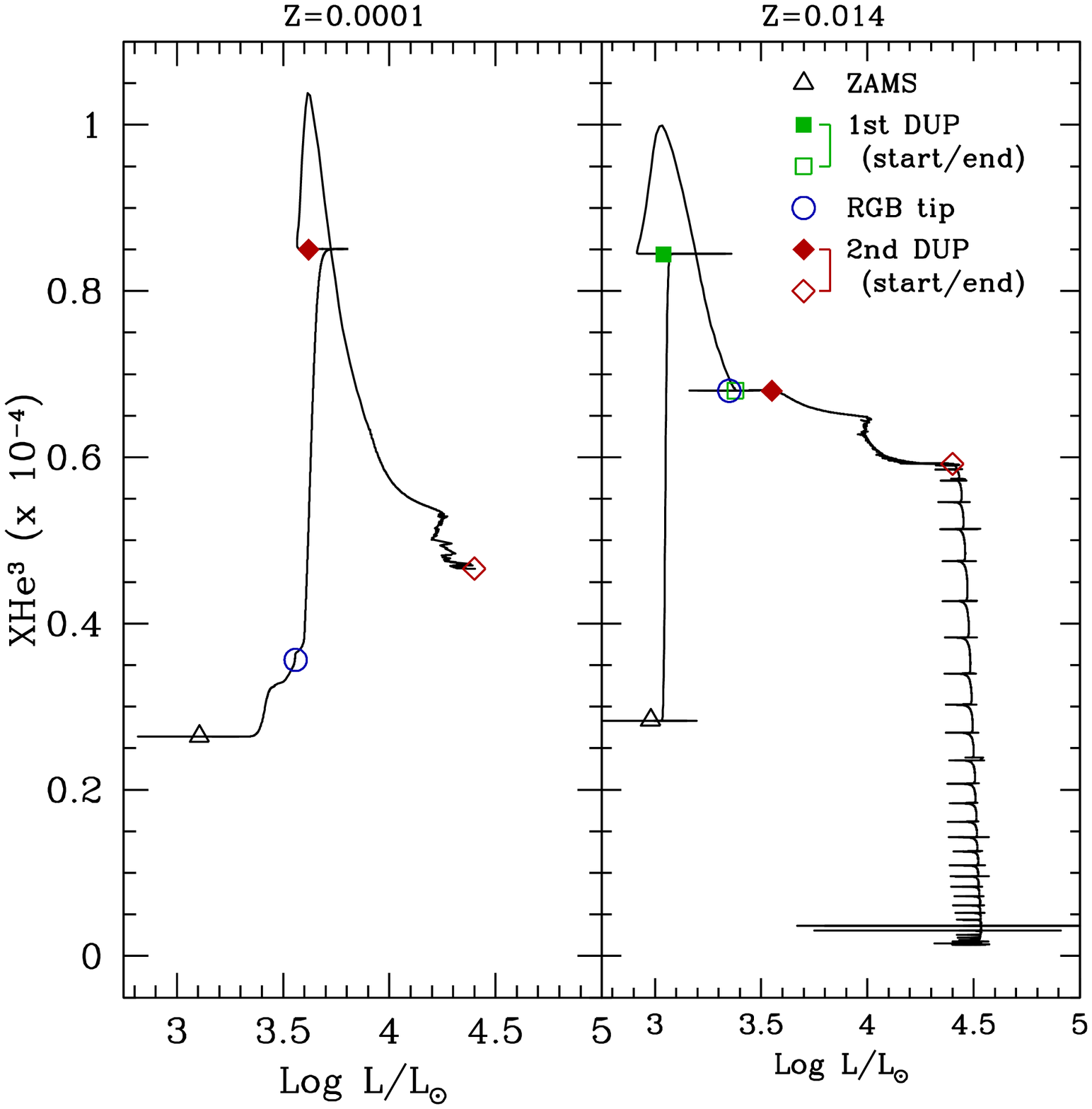}
		\includegraphics[angle=0,width=9cm]{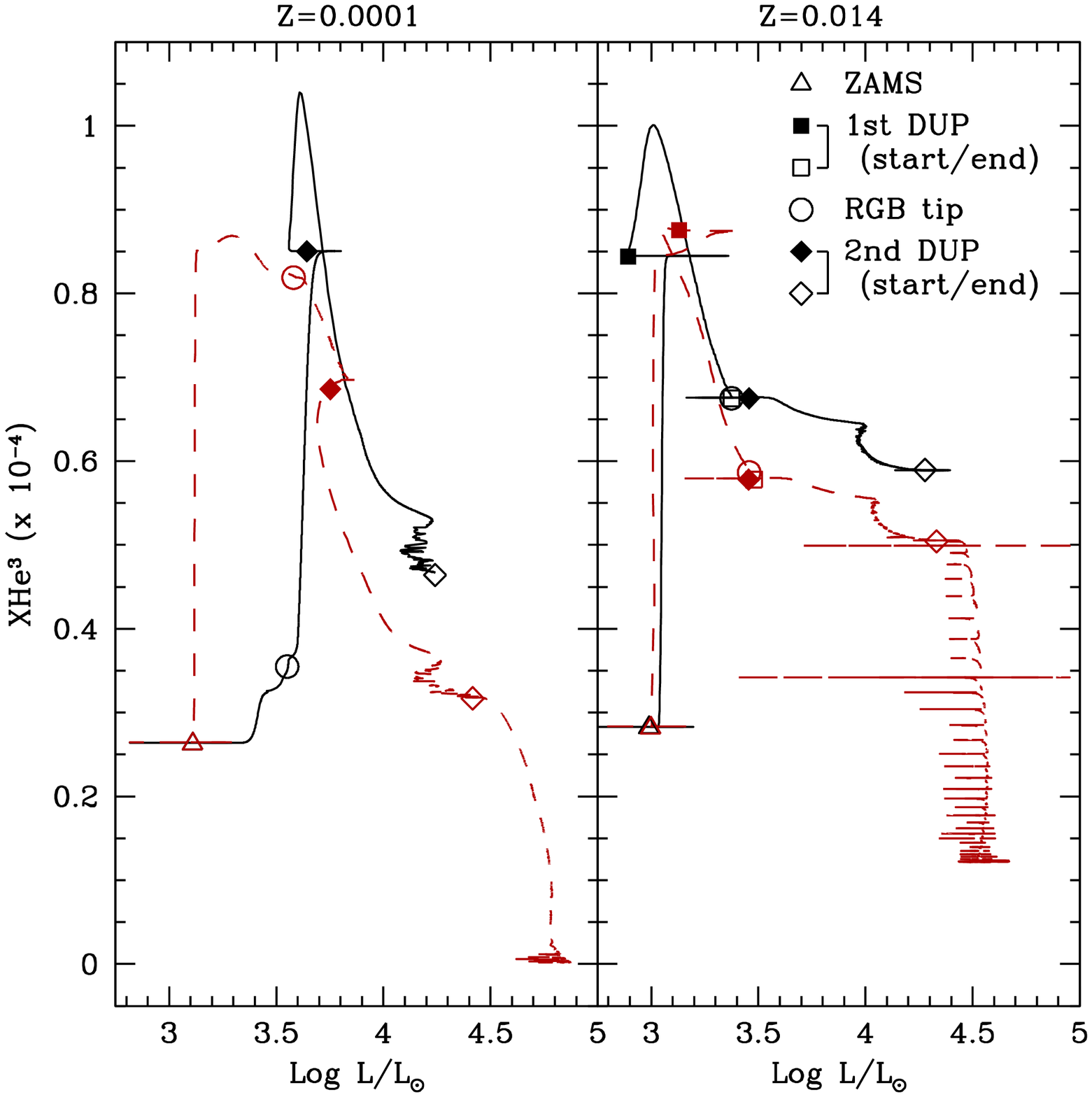}
	  \caption{Same as Fig.~\ref{fig:He3_1p5_0001_014} for 6~M$_{\odot}$ models (but with different ordinate). {\sl (Left)} Standard case. The low-metallicity model is computed up to the end of second dredge-up only, while the Z$_{\odot}$ one is carried out until the AGB tip.
	   {\sl (Right)} Models including thermohaline mixing only (black solid lines), and including both thermohaline and rotation-induced mixings (red dashed lines; $V_{ZAMS}/V_{crit}=0.45$). These rotating models are computed until the AGB tip.}
	\label{fig:He3_6_0001_014}
\end{figure*}

\begin{figure*} 
	\centering
	~\hfill \hfill Standard models\hfill\hfill\hfill Thermohaline +Rotating models\hfill ~ \\
		\includegraphics[angle=0,width=9cm]{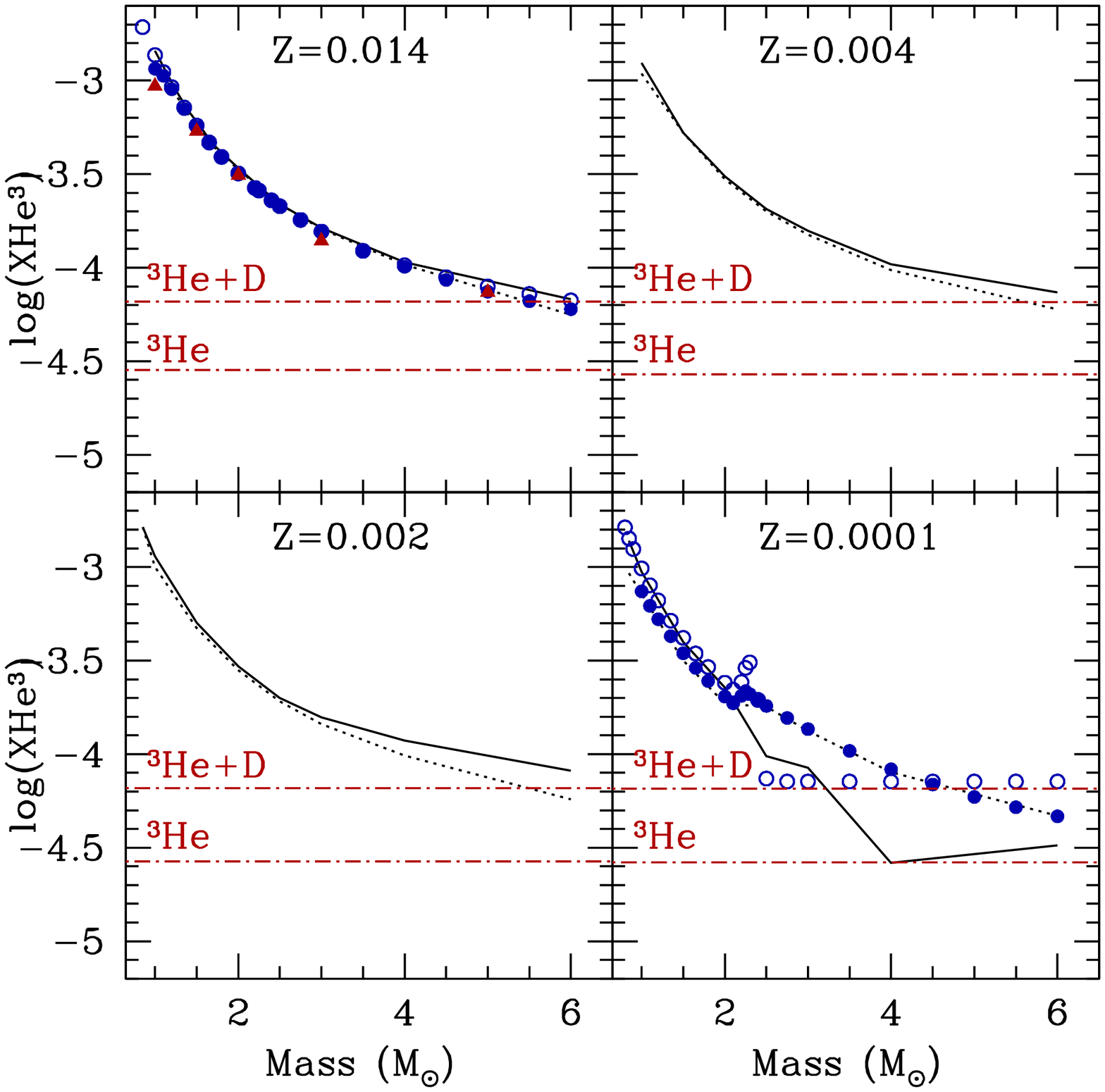}
		\includegraphics[angle=0,width=9cm]{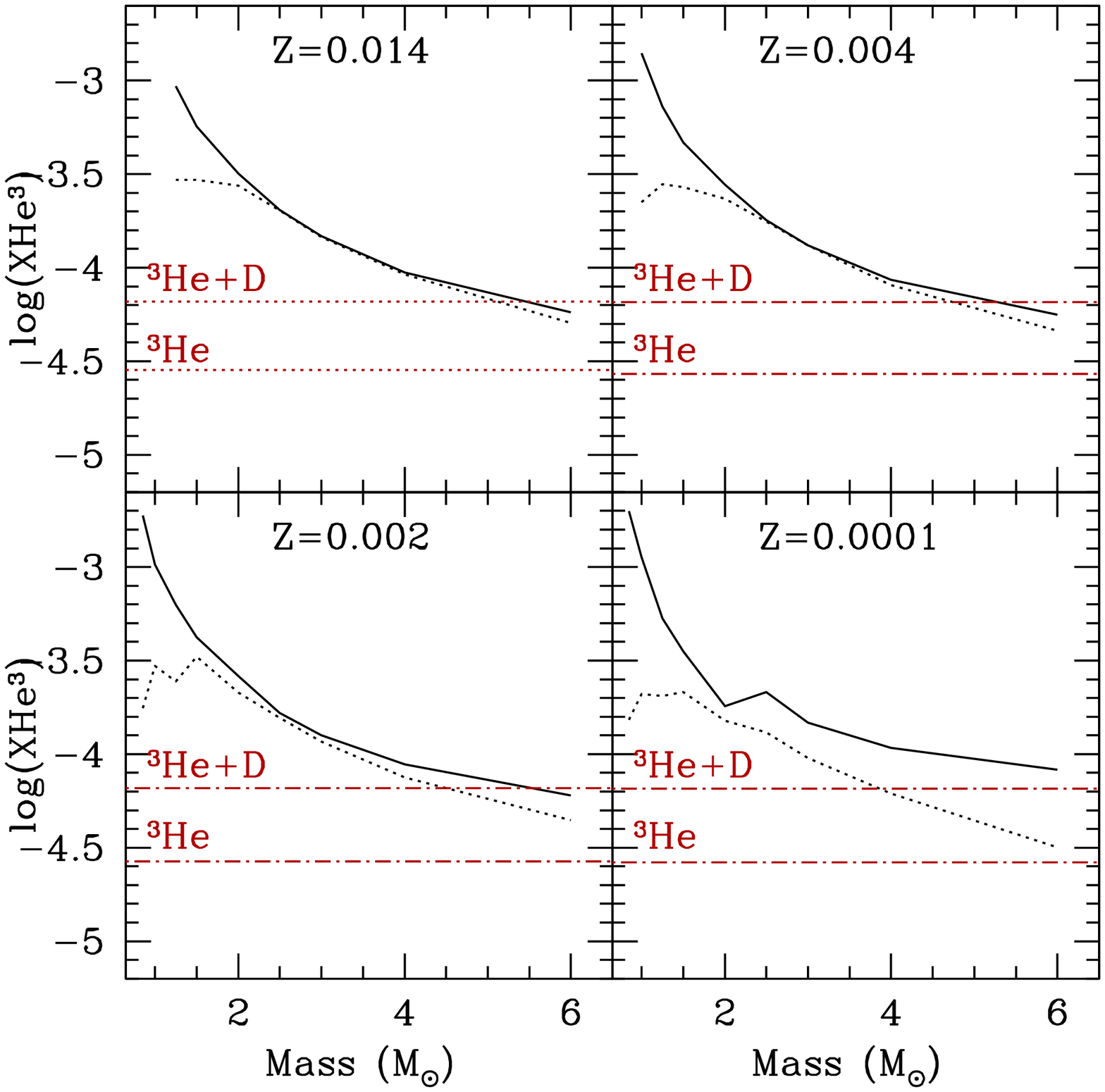}
	  \caption{Surface abundance of $^{3}$He (in mass fraction) at the end of first and second dredge-up episodes (solid and dashed black lines respectively) as a function of initial stellar mass and for the four metallicities considered (for the 3 to 6~M$_{\odot}$ standard models at Z=0.0001 that do not undergo first dredge-up, we show instead the $^{3}$He surface abundance at the main-sequence turnoff). Red dotted curves indicate the initial $^{3}$He abundance and total D$+^{3}$He assumed at stellar birth. {\sl (Left)} Standard models.  
	  Full red triangles correspond to the predictions by \cite{Weissetal96} at the end of 1DUP for Z${_\odot} $; blue circles (open and full at the end of 1DUP and 2DUP respectively) are predictions by  \cite{SaBo99} for Z${_\odot} $ and Z=$10^{-4}$. {\sl (Right)} Models including thermohaline and rotation-induced mixings. 
	}
	\label{fig:3He1st2dDUPstandard}
\end{figure*}

When stars move towards the RGB, their convective envelope deepens until they reach the region indicated by the vertical lines in Fig.\ref{fig:3HepeakvsMvsZ}, and engulfs all or part of the fresh $^3$He peak. The first dredge-up efficiency (in terms of maximum penetration depth of the convective envelope) decreases with decreasing metallicity as shown in Fig.\ref{fig:Bottom_CE} (at Z=$10^{-4}$, no first dredge-up occurs for stars more massive than $\sim$ 3.0~M$_{\odot}$).
As a consequence, the surface abundance of $^3$He increases by a factor that depends on the initial stellar mass as well as on the metallicity,
as can be seen in Fig.\ref{fig:He3_1p5_0001_014}, \ref{fig:He3_6_0001_014}, and \ref{fig:3He1st2dDUPstandard}\footnote{The evolution of the $^3$He abundance at the surface of 6~M$_{\odot}$ standard models depicted in Fig.~\ref{fig:He3_6_0001_014} for the two extreme Z values shows a couple of peculiarities compared to the case of low-mass stars. For such a relatively massive star, the base of the convective envelope withdraws very quickly on the pre-main sequence, precluding any increase of the surface $^3$He during that phase. However, because of mass loss, the layers enriched in $^3$He by pristine D-burning appear at the surface later on. In addition, no first dredge-up occurs in the lowest Z 6~M$_{\odot}$ model (see also Fig.~\ref{fig:3He1st2dDUPstandard}).}.

During the subsequent evolution on the RGB that proceeds on shorter timescales compared to the main-sequence lifetime, H-burning is concentrated in a very small (both in mass and radius) radiative shell that surrounds the degenerate helium core and is dominated by the CNO-cycle.  
There is thus no significant further $^3$He production inside evolved stars. 
In addition, the temperature in their convective envelope remains always low enough to preserve the freshly dredged-up $^3$He.
Consequently and as far as standard models are concerned,  there is no other change in the $^3$He surface abundance until the stars undergo the second dredge-up on the early-AGB.  During this short episode the convective envelope deepens again (see Fig.~\ref{fig:Bottom_CE}) 
and reaches $^3$He-free regions, which induces a slight decrease of the surface abundance of this element (see Fig.\ref{fig:He3_1p5_0001_014}, \ref{fig:He3_6_0001_014}, and \ref{fig:3He1st2dDUPstandard}). 

Except in the few cases discussed below (see Tables 1 to 4), we stopped our standard computations at that phase. For low-mass stars the standard models published in the literature predict that the $^3$He stored in the stellar convective envelope survives the TP-AGB phase before it is injected into the ISM by stellar wind and planetary nebula ejection (see references in \S~\ref{standardliterature}). This agrees with our standard [1.5, 2$M_{\odot}$; Z$_{\odot}$], [1.0; Z=0.004], and [1.0, 1.5$M_{\odot}$; Z=0.002] models that we computed up to the AGB tip and for which surface $^{3}$He abundance remains constant during the TP-AGB phase.
Note, however, that in stars with initial masses higher than $\sim$ 3.5-4~M$_{\odot}$, hot-bottom burning during the TP-AGB phase induces $^3$He-burning at the base of the convective envelope and thus reduces the abundance of this element in the whole envelope and at the stellar surface \citep{SaBo92,Weissetal96, FoCh97,SaBo99}. Because of this process $^3$He can even be completely destroyed at the AGB tip in the most massive intermediate-mass stars (see the standard [6~M$_{\odot}$; Z$_{\odot}$] model in Fig.~\ref{fig:He3_6_0001_014}), so that the corresponding net yields are negative. This will be discussed in more detail in \S \ref{thermoharotationmodels}, where we present the computations up to the TP-AGB tip for complete models including the effects of thermohaline and rotation-induced mixing. 

\subsubsection{Surface abundances prior to the TP-AGB and comparison with standard models from the litterature\label{standardliterature}}

In Fig.\ref{fig:3He1st2dDUPstandard} we present the standard predictions for the surface abundance of $^3$He after both first and second dredge-up episodes (solid and dashed lines respectively) as a function of initial stellar mass for the four considered metallicity values. 
Note that the 3 to 6~M$_{\odot}$ models at Z=0.0001do not undergo the first dredge-up (see Fig.~\ref{fig:Bottom_CE}). In this case we show the $^3$He surface abundance at the end of the main sequence and after second dredge-up (solid and dashed).

The assumed initial $^{3}$He abundance as well as the total initial D$+^{3}$He are also indicated in Fig.\ref{fig:3He1st2dDUPstandard}. 
This confirms that in the standard case, stars with an initial mass lower than $\sim$ 3.5-4~M$_{\odot}$ that do not undergo hot-bottom burning are expected to strongly enrich the Universe with $^{3}$He. The mass dependency described above also clearly shows up. 

In this figure we also compare our standard predictions with those by \cite{Weissetal96} and \cite{SaBo99} at solar metallicity on one hand, and those by  \cite{SaBo99} for Z=$10^{-4}$ on the other hand. An excellent agreement is found with these standard theoretical models.

\subsection{Models including rotation-induced mixing and thermohaline instability \label{thermoharotationmodels}}

In Paper I we discussed at length the impact of both rotation-induced mixing and thermohaline instability on the structure and chemical properties of low- and intermediate-mass stars at various phases of their evolution.
Here we only briefly summarize the main points, focusing on $^3$He. 

Note that in low-mass stars, thermohaline mixing induced by the molecular-weight inversion due to the $^3$He($^3$He,2p)$^4$He reaction sets in only at the RGB bump. 
Up  to that phase, predictions for models including only this process (i.e., not including rotation-induced effects) are therefore similar to the standard ones described above (in particular the predictions are the same for $^3$He surface values caused by first dredge-up), and they start differing  only on the upper RGB. In the case of intermediate-mass stars, thermohaline mixing starts playing a role even later, i.e., on the TP-AGB phase.
However, rotation-induced mixing has an impact already in the earlier phases  for all stellar masses, as described below.

\subsubsection{Main-sequence abundance profiles and first dredge-up}

As known for a long time (see references in Paper I), rotation-induced mixing modifies the internal structure of main-sequence stars, and smoothes out the abundance gradients with respect to the standard case. 
Consequently, the $^3$He production is moderatly affected by the slightly higher temperature in the rotating models compared to the standard case for a given stellar mass and metallicity during central H-burning. 
But more importantly, the $^3$He peak is spread out and fresh $^3$He is expected to reach the stellar surface during the main-sequence lifetime. The resulting profiles at turnoff are shown in Fig.\ref{fig:3HepeakvsMvsZ} and can be compared to the standard ones (left and right panels respectively). 
At that evolution point, the total $^3$He content for a star of given initial stellar mass and metallicity is slightly lower in the rotating case. 
As can be seen in Fig.\ref{fig:3HepeakvsMvsZ}, the higher the initial rotation velocity, the stronger the effect.

Additionally, the structural and chemical changes caused by rotation on the main sequence favour a slightly deeper penetration of the convective envelope during the subsequent first dredÄge-up episode (because the mass of the He-core is slightly larger at the end of the main sequence in the rotating case), which leads to the engulfment of larger $^3$He-free regions that lie below the peak. As a consequence, post dredge-up $^3$He values are lower when rotation is accounted for than in the standard models, as can be seen in Fig.\ref{fig:He3_1p5_0001_014}, \ref{fig:He3_6_0001_014},  and \ref{fig:3He1st2dDUPstandard}, and in Tables~1 to 4.
The effects are stronger for increasing stellar mass and decreasing metallicity.

\subsubsection{Red giant branch}

\begin{figure}[htbf] 
	\centering
		\includegraphics[angle=0,width=9cm]{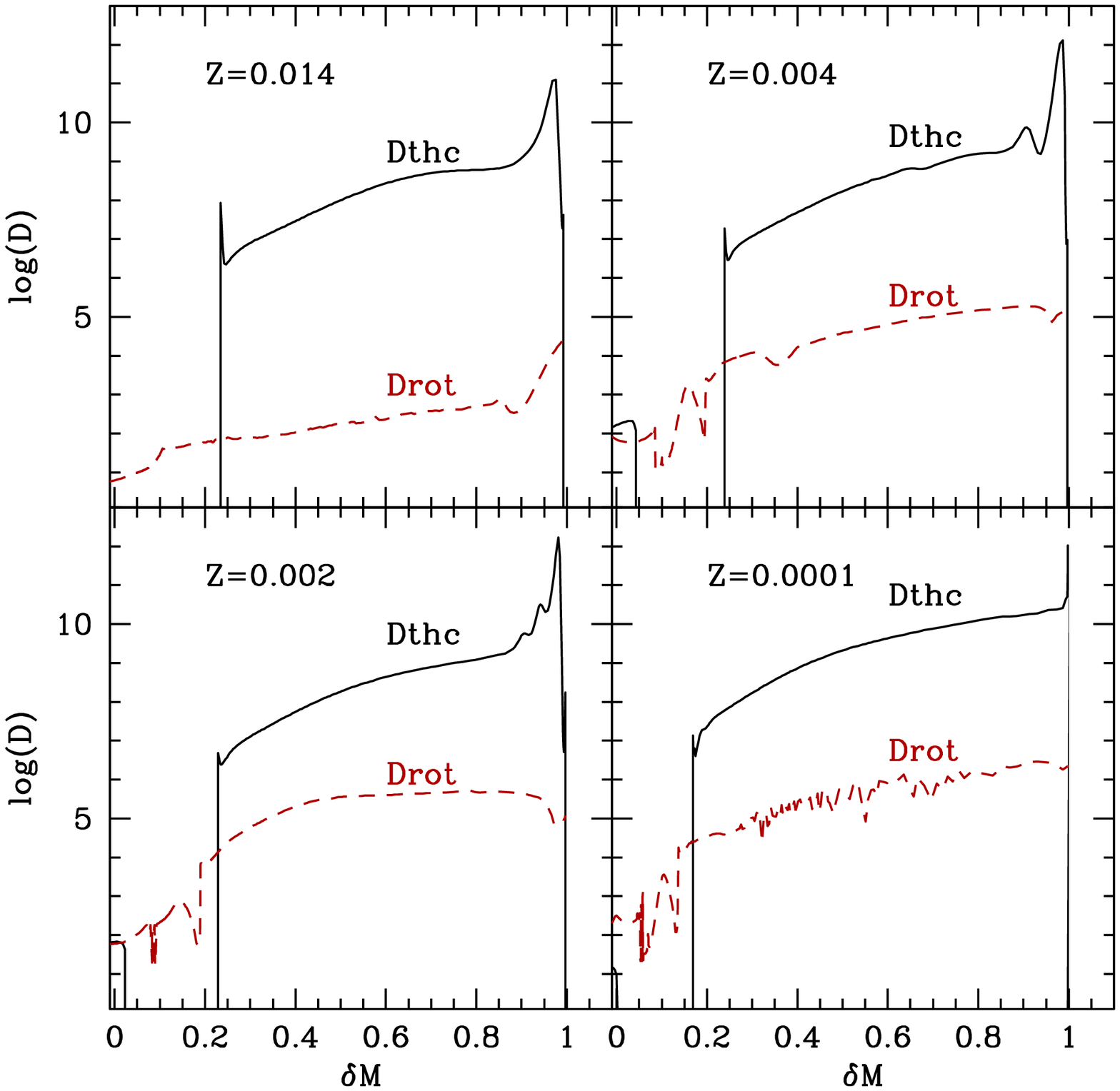}
	  \caption{Thermohaline diffusion coefficient, D$_{\rm thc}$, and total rotation diffusion coefficient,  D$_{\rm rot}$, (solid and dashed lines respectively) as a function of reduced stellar mass, $\delta {\rm M}$ ($\delta$M = $\frac{{\rm M}_r - {\rm M}_{\rm HBS}}{{\rm M}_{\rm BCE} - {\rm M}_{\rm HBS}}$ = 0 at the base of the hydrogen-burning shell, M$_{\rm HBS}$; 
	   $\delta$M  = 1 at the base of the convective envelope, M$_{\rm BCE}$) for 1.25M$_{\odot}$ models at different metallicities. In each case the evolution point is chosen at the luminosity when thermohaline mixing connects the hydrogen-burning shell with the convective envelope (see text)}
	\label{fig:Dthermo_Drot_1p25_Z}
\end{figure}

In the advanced evolution phases the total diffusion coefficient associated to rotation is too low to induce abundance changes at the stellar surface \citep[see below, and also ][ and Paper I]{Chaname05,Palacios06,CaLa08,CantielloLanger2010}.
However, for low-mass stars the surface abundances change after the RGB bump with respect to the post dredge-up values, because of thermohaline mixing induced by the molecular weight inversion created by the $^3$He($^3$He,2p)$^4$He reaction in the outer wing of the hydrogen-burning shell. 
Figure \ref{fig:Dthermo_Drot_1p25_Z} compares the diffusion coefficient associated to the thermohaline instability, D$_{\rm thc}$, for a 1.25~M$_{\odot}$ star at different metallicities, to the total diffusion coefficient associated to rotation,  D$_{\rm rot}$, that characterizes the transport of chemicals caused by meridional circulation and shear turbulence. For each model these quantities are shown at the evolution point on the RGB when the surface abundances start changing because of thermohaline mixing (this refers to the evolution point C$_{1.25}$ in Fig.1 of Paper I) . 
In all cases, D$_{\rm thc}$ is at least 2 to 3 orders of magnitude higher than D$_{\rm rot}$ \citep[see also Paper I, and][]{CaLa08, CantielloLanger2010}.

As discussed in detail in Paper I and in \cite{ChaZah07a}, the present prescription for the thermohaline diffusivity accounts for the observed behaviour of $^{12}$C/$^{13}$C, [N/C], and lithium in low-mass stars that are more luminous than the RGB bump. It simultaneously leads to strong $^3$He depletion in the stellar envelope, as can be seen in Fig. \ref{fig:He3_1p5_0001_014} \citep[see also e.g.][]{ChaZah07a}.
However, $^3$He is not completely destroyed at the RGB tip, and therefore it can drive the thermohaline instability in the latter evolution phases, as discussed below.

\subsubsection{Early-AGB and $^3$He surface abundances after the second dredge-up}

\begin{figure*} 
	\centering
		\includegraphics[angle=0,width=9cm]{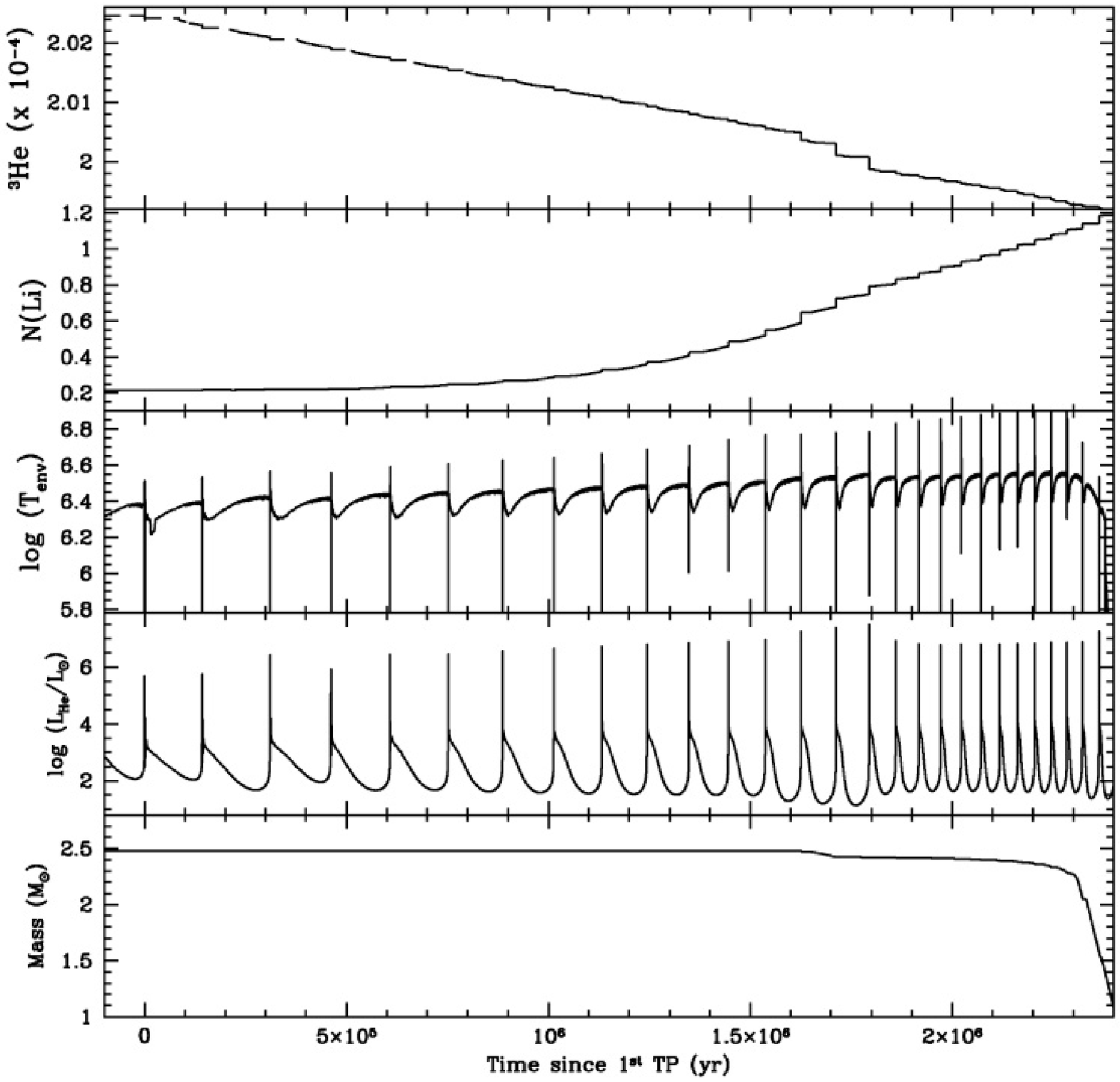}
		\includegraphics[angle=0,width=9cm]{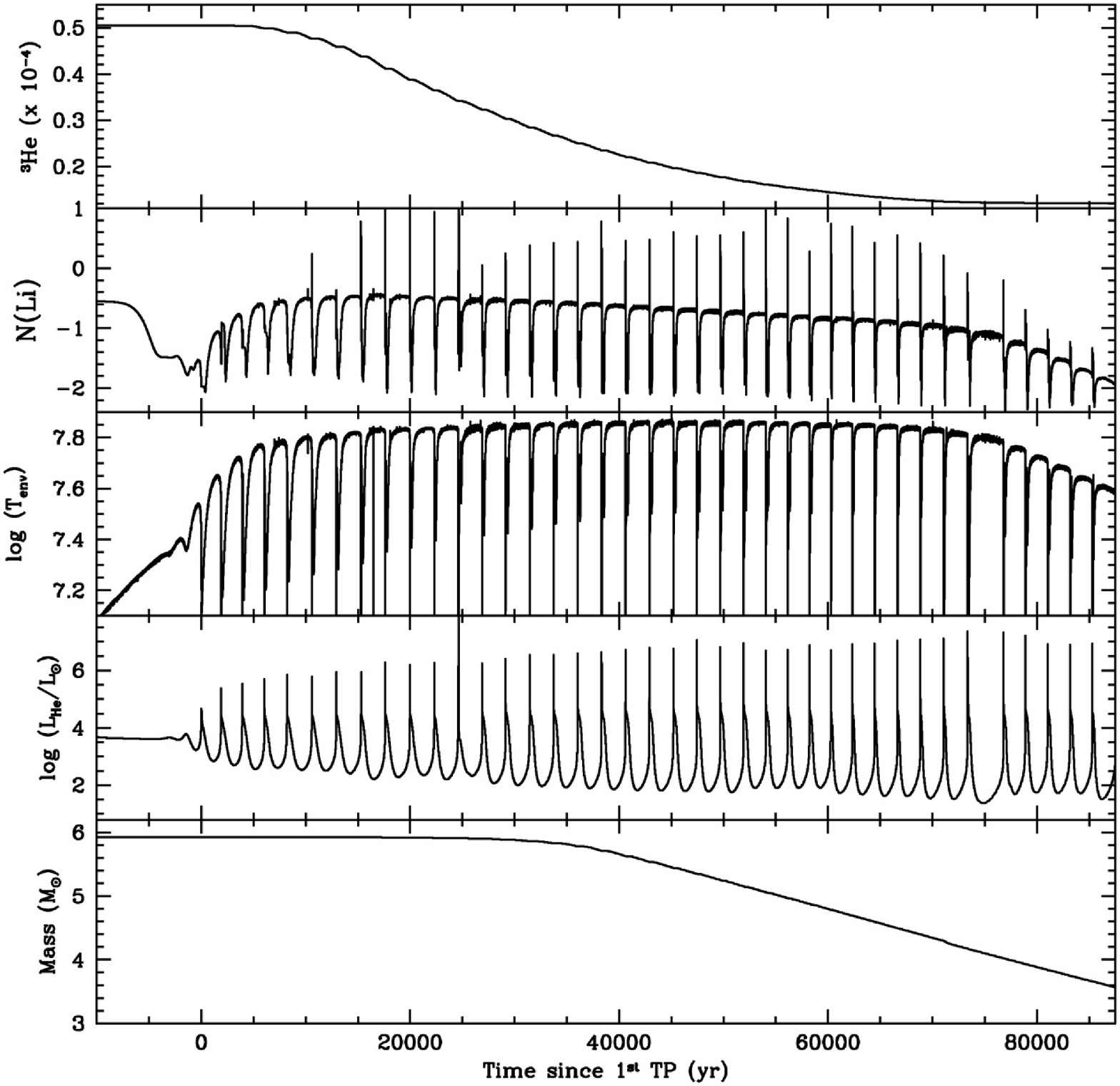}
	  \caption{{\sl (From top to bottom)} Evolution along the TP-AGB of the surface abundance of $^{3}$He, of N(Li), of the temperature at the base of convective envelope, of the helium-burning luminosity, and of the total stellar mass. The abscissa is the time since the first thermal pulse. {\sl Left and right respectively:}  2.5 and 6~$M_{\odot}$ at $Z_{\odot}$ models computed with thermohaline and rotation-induced mixings}
	\label{fig:TPs_2p5_014_He3_Li_LHe}
\end{figure*}

In stars with initial masses lower than, or equal to 2 and 1.5~M$_{\odot}$ for Z=Z$_{\odot}$ and Z=0.0001 respectively (with intermediate values for the upper mass limit for the intermediate metallicities), the $^3$He abundance continues to decrease at the stellar surface between the end of central helium burning and the second dredge-up episode through thermohaline mixing, as can be seen in Fig.\ref{fig:He3_1p5_0001_014}.
On the other hand, in more massive stars rotation induces a decrease of the $^3$He surface abundance during central helium burning and on the early-AGB (see Fig.\ref{fig:He3_6_0001_014}) before they undergo the second dredge-up (Fig.\ref{fig:Bottom_CE}). 
 When integrated over the whole evolution, changing V$_{\rm ZAMS}$/V$_{\rm crit}$ from 0.45 to $\sim$ 0.9 leads to a decrease of the $^3$He surface abundance after the second dredge-up of $\sim$ 5, 11.5, and 22, 23 $\%$  in the [2~M$_{\odot}$; Z$_{\odot}$], [4~M$_{\odot}$; Z$_{\odot}$], and [2~M$_{\odot}$; Z$=10^{-04}$], [4~M$_{\odot}$; Z$=10^{-04}$] models respectively (see Table~\ref{tableyields014}).

The resulting surface $^3$He values after second dredge-up are shown in Fig.\ref{fig:3He1st2dDUPstandard} (right panels) for the non-standard models over the whole mass and metallicity range.
In summary, compared to the standard case (left panels), one finds that the thermohaline instability dominates in reducing the $^3$He content of low-mass stars, while the dominating process is rotation for intermediate-mass stars. The impact of both mechanisms increases with decreasing metallicity.

\begin{figure} 
	\centering
		\includegraphics[angle=0,width=9cm]{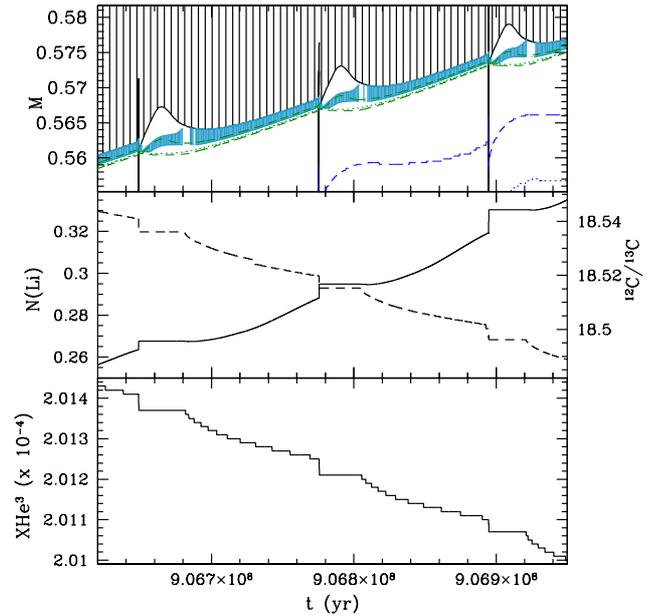}
	  \caption{Zoom on the effect of thermohaline mixing between the  thermal pulses number 7 and 9 for the 2.5~$M_{\odot}$ rotating model at $Z_{\odot}$. {\sl (Bottom)} Kippenhahn diagram. The hatched zone is the convective envelope; the blue shaded region corresponds to the layers where the thermohaline instability develops; the green and blue dashed lines surround the hydrogen and the helium burning shell respectively (in both cases, the dotted lines show the region of maximum nuclear energy production). {\sl (Middle)} Evolution of the surface $^7$Li abundance (full line) and $^{12}$C/$^{13}$C ratio (dashed line). {\sl (Bottom)} Evolution of the surface $^3$He abundance (in mass fraction)
	  }
	\label{fig:3DUP_time_Li_He3_c1213_2p5_014_throt}
\end{figure}

\subsubsection{TP-AGB}

After the second dredge-up, $^3$He remains in sufficient quantity to drive thermohaline mixing during the TP-AGB in stars that do not undergo hot-bottom burning (i.e., with initial mass below $\sim$ 3~M$_{\odot}$).  
As discussed in Paper I \citep[see also ][]{CantielloLanger2010,Stancliffe10}, this leads to even greater (although modest) depletion of $^3$He associated to $^7$Li production during that phase. This is depicted in Fig.\ref{fig:TPs_2p5_014_He3_Li_LHe}  as a function of time since the first thermal pulse for a rotating [2.5~M$_{\odot}$; Z$_{\odot}$] model computed with thermohaline mixing up to the AGB tip (left panels). This figure also shows the temperature at the base of the convective envelope, the helium-burning luminosity, and the total stellar mass.
When compared to the [1.25~M$_{\odot}$; Z$_{\odot}$] model discussed in Fig.7 of Paper I, a slight difference shows up. In the 2.5~M$_{\odot}$ and from pulse number four on, the convective envelope deepens immediatly after each pulse inside the thermohaline region, re-inforcing the thermohaline effect modifying the surface abundances, as can be seen in Fig.\ref{fig:3DUP_time_Li_He3_c1213_2p5_014_throt} (which focusses on pulses number 7 to 9).
Note, however, that the decrease of $^3$He at the stellar surface owing to the whole process along the TP-AGB is modest, and that $^3$He is not completely destroyed when these stars reach the TP-AGB tip (see also Tables 1 to 3). In this framework low-mass stars accordingly remain net $^3$He producers.

Figure \ref{fig:TPs_2p5_014_He3_Li_LHe} also shows the evolution during TP-AGB of $^{3}$He and $^7$Li abundances at the  surface of the non-standard [6.0M$_{\odot}$, Z$_{\odot}$] model. 
For such a relatively massive star $^{3}$He is not abundant enough to drive thermohaline mixing during thermal pulses.
However, the temperature at the base of the convective envelope is sufficient to engage hot-bottom burning.  
As a consequence, the surface abundance of $^{3}$He decreases very rapidly, until complete destruction. The simultaneous $^{7}$Li enrichment at the stellar surface is caused in this case by the \cite{CameronFowler1971} process.

\section{Yields of $^{3}$He and conclusions \label{yields}}

\begin{figure*} 
	\centering
		\includegraphics[angle=0,width=9cm]{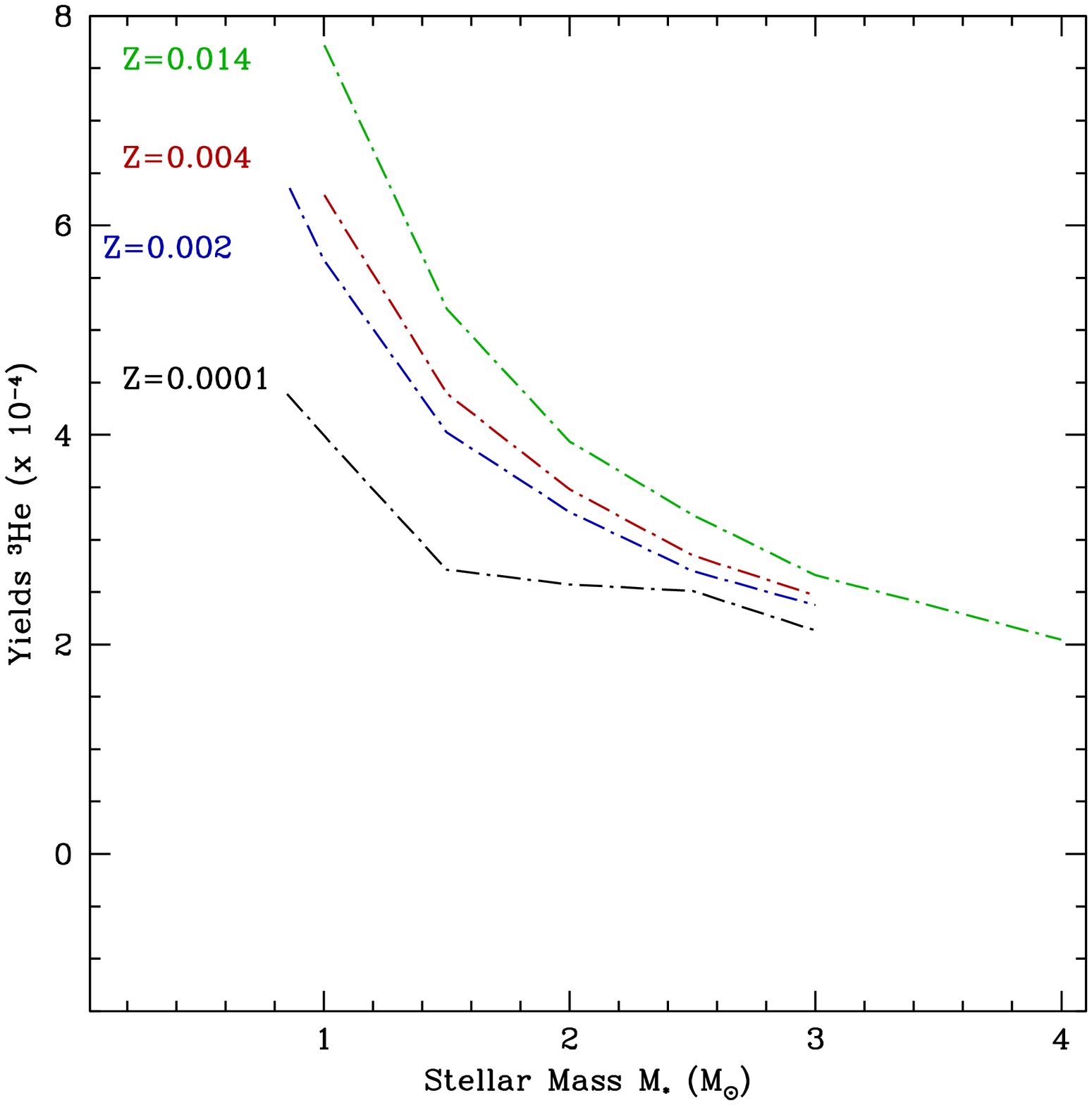}
				\includegraphics[angle=0,width=9cm]{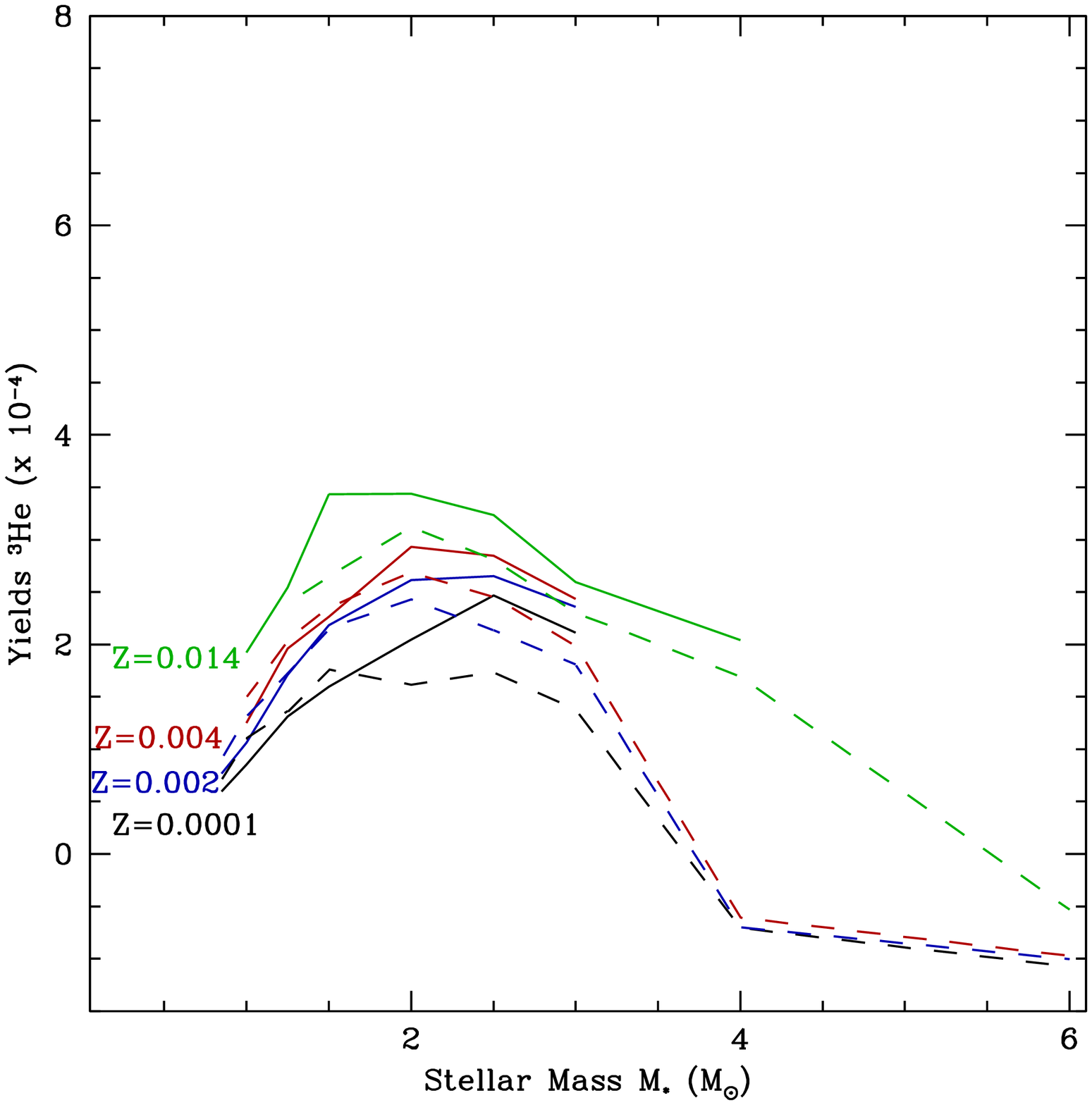}
	  \caption{Yields of $^{3}$He as a function of stellar mass, at different metallicities. {\sl (Left)} Predictions for standard models.  {\sl (Right)} Predictions including thermohaline mixing only (solid lines) and including both thermohaline and rotation-induced mixings (dashed lines, for models computed up to the AGB tip or to $^{3}He$ exhaustion).}
	\label{fig:yieldsstd}
\end{figure*}

We can now summarize our study by quantifying in terms of yields the impact of thermohaline instability and rotation-induced mixing on the production and destruction of $^{3}$He by low- and intermediate-mass stars at various metallicities. 
The net yields of $^{3}$He are shown in Figure \ref{fig:yieldsstd} for all our models and are also given in Tables~1 to 4. There one can find for selected models comparisons between the yields obtained when computing the models until the end of second dredge-up only (case A), or up to the AGB tip (case C). Because the decrease of $^{3}$He caused by thermohaline mixing during the TP-AGB is very modest for low-mass stars  compared to what happens on the RGB, the difference between the extrapolated yields (A) and the ones obtained from the full computations (C) is below $\sim$ 12 $\%$ only.

\noindent Our main results may be summarized as follows:
\begin{itemize}
\item Over the whole mass and metallicity range considered,  the total $^{3}$He content is lowered when rotation-induced mixing is accounted for compared to the standard case.
\item For low-mass stars (M$<$2-2.2 M$_{\odot}$) that produce large quantities of this light element through the pp-chains on the main sequence, thermohaline mixing on both the RGB and the TP-AGB is dominant in reducing the final $^{3}$He yield. These stars remain net producers of $^{3}$He however, although their contribution to the Galactic evolution of this light element is strongly reduced compared to the standard framework.
\item For intermediate-mass stars thermohaline mixing does not operate on the shorter RGB (these stars ignite central He-burning before reaching the RGB bump). 
\item For those with masses between 2-2.2 and 3-4~M$_{\odot}$, thermohaline mixing 
leads however to modest $^{3}$He depletion during the TP-AGB phase, associated with lithium production.
\item In more massive intermediate-mass stars, $^{3}$He is strongly reduced through the action of rotation that lowers the upper mass limit for stars that are net $^{3}$He producers, and is additionally destroyed through hot-bottom burning while they climb the TP-AGB. In these objects Li is produced through the classical Cameron-Fowler mechanism.
\end{itemize}
The impact on Galactic evolution predictions  of these new $^{3}$He yields will be presented in a forthcoming paper.

%=============================table fraction de masse de He3 et Yields ˆ Zsun ==========================
\begin{table*}
	\hspace{3cm}
	\caption{Model results for metallicity of Z=0.0001.
	} 
         
         \begin{threeparttable}
         \centering
                                  
	\begin{tabular}{| c | c | c | c | c | c | c || c | c |}  
	\hline  
	&                         &                        &                          &                           & \multicolumn{2}{c||}{Mass fraction $^{3}$He  $^{[3]}$}& &\\
	M    &                 & V$_{\rm ZAMS}$/V$_{crit}$ $^{[1]}$ & V$_{\rm ZAMS}$ $^{[1]}$ & life time at TO$^{[2]}$ & 1DUP & 2DUP & Yield $^{3}$He $^{[4]}$& $^{[5]}$  \\
	(M$_{\odot}$) & &                      &        (km.sec$^{-1}$)     &    (yr)                 &              &             & (M$_{\odot}$)             & \\
	\hline \hline
	0.85 & stand.      & - & -      & $1.06.10^{10}$ & $1.38.10^{-03}$ & $9.25.10^{-04}$ & $4.39.10^{-04}$  & A \\
		& thermoh  & - &-       & $1.06.10^{10}$ & $1.39.10^{-03}$ & $1.40.10^{-04}$  & $5.95.10^{-05}$ &  A \\
		& th. +rot.   & 0.45 &115 & $1.08.10^{10}$ & $1.79.10^{-03}$  & $1.56.10^{-04}$ & $7.17.10^{-05}$  &A  \\
		& th. +rot. (K=$10^{31}$)  & 0.45 &115 & $1.08.10^{10}$ & $1.99.10^{-03}$  & $1.57.10^{-04}$  & $6.89.10^{-05}$ &A   \\
          \hline
	1.0   & stand.  & - & - & $5.93.10^{09}$ &  $9.38.10^{-04}$&  $7.03.10^{-04}$ & $3.99.10^{-04}$ & A \\
		& thermoh &- & - & $5.93.10^{09}$ & $9.38.10^{-04}$ & $1.67.10^{-04}$ & $8.54.10^{-05}$ &A  \\
	         & th. +rot.  & 0.45 &116  & $6.06.10^{09}$ &  $1.13.10^{-03}$& $2.06.10^{-04}$ & $1.10.10^{-04}$  &A   \\
          \hline
	1.25 & thermoh & - &- & $2.70.10^{09}$ & $5.74.10^{-04}$ & $1.95.10^{-04}$ & $1.31.10^{-04}$  &A  \\
	         & th. +rot.  & 0.45 &125  & $2.65.10^{09}$& $5.30.10^{-04}$  & $2.02.10^{-04}$ & $1.36.10^{-04}$ & A\\
          \hline
	1.5   & stand.  & - & - & $1.46.10^{09}$ & $3.98.10^{-04}$ & $3.16.10^{-04}$ &  $2.71.10^{-04}$  & A \\
		& thermoh & - & -  & $1.48.10^{09}$ & $3.98.10^{-04}$ &$1.95.10^{-04}$ & $1.60.10^{-04}$   & A \\
	         & th. +rot.  & 0.45 &134 & $1.52.10^{09}$ &$3.53.10^{-04}$ & $2.13.10^{-04}$ & $1.76.10^{-04}$  & A   \\
          \hline
          2.0  & stand.     & - & -  & $5.70.10^{08}$ & $2.26.10^{-04}$ & $1.85.10^{-04}$ & $2.57.10^{-04}$  &A   \\
                  & thermoh & - &- & $5.69.10^{08}$ &$2.26.10^{-04}$  &$1.85.10^{-04}$  & $2.05.10^{-04}$  & A  \\
                  & th. +rot.  & 0.45 &150  & $6.08.10^{08}$ &$1.81.10^{-04}$  & $1.51.10^{-04}$ & $1.61.10^{-04}$  &A   \\
                  & th. +rot.  & 0.90  & 300   & $6.08.10^{08}$  & $1.39.10^{-04}$   & $1.17.10^{-04}$  & $1.18.10^{-04}$   & A    \\
          \hline
         2.5   & stand.  & - &- & $3.08.10^{08}$ & $9.73.10^{-05}$&$1.79.10^{-04}$ & $2.50.10^{-04}$   & A  \\
         		& thermoh & - &  - & $3.08.10^{08}$ & $9.63.10^{-05}$ & $1.76.10^{-04}$ & $2.47.10^{-04}$   & A  \\
                  & th. +rot.  & 0.45 & 162  & $3.24.10^{08}$  & $2.13.10^{-04}$  &  $1.31.10^{-04}$ & $1.73.10^{-04}$ & A\\
          \hline
         3.0   & stand.  & - & - & $2.06.10^{08}$ & $8.46.10^{-05}$ & $1.33.10^{-04}$  & $2.13.10^{-04}$ &A  \\
         		& thermoh & - & - &  $2.06.10^{08}$ & $8.46.10^{-05}$ & $1.32.10^{-04}$ & $2.11.10^{-04}$ & A\\
                  & th. +rot.  & 0.45 & 170  & $2.15.10^{08}$&  $1.47.10^{-04}$ & $9.51.10^{-05}$ & $1.37.10^{-04}$  &  A  \\
         \hline 
         4.0 & stand.  & - & - & $1.08.10^{08}$ &  $2.63.10^{-05}$& $8.01.10^{-05}$  & - &A \\
         		& thermoh & - & - &  $1.08.10^{08}$ & $2.63.10^{-05}$& $8.09.10^{-05}$ & - &A \\
                  & th. +rot.  & 0.45 & 152 & $1.13.10^{08}$ & $1.08.10^{-04}$ & $6.18.10^{-05}$ &  $-7.1.10^{-05}$  & B\\
                  &th.+rot. & 0.90 & 304 & $1.18.10^{08}$& $9.48.10^{-05}$ & $4.90.10^{-05}$ & - & A \\
         \hline
         6.0  & stand.  & - & - & $5.15.10^{07}$ & $3.26.10^{-05}$ & $4.65.10^{-05}$  & - &A \\
         		& thermoh & - & - &  $5.16.10^{07}$ & $3.26.10^{-05}$ & $4.68.10^{-05}$ & -&A\\
                 & th. +rot.  & 0.45 & 175  & $5.35.10^{07}$ &  $8.26.10^{-05}$& $3.18.10^{-05}$ & $-1.08.10^{-4}$  & B\\
         \hline
	\end{tabular}
	\label{tableyields0001}
	
        \begin{tablenotes}
        \item Each row contains entries for different assumptions: standard (without thermohaline or rotation-induced mixing); thermohaline mixing only; thermohaline and rotation-induced mixing. \\
        \item[1] The initial rotation on the ZAMS 
        \item[2] Life time at turn-off
        \item[3] Mass fraction of $^{3}$He at the stellar surface after first and second dredge-up
        \item[4] Yields of $^{3}$He
        \item[5] The phase where the computations are stopped is given in the last column: Case A: Model computed until the end of the second dredge-up on the early-AGB.  Case B: Model computed along the TP-AGB until the mass fraction of $^3$He at the surface is below $\sim$10$^{-5}$ due to the HBB process. Case C: Model computed until the end of the superwind phase. In that former case we also give the extrapolated A yield value (in brackets, A)
        
       	\end{tablenotes}
	
	\end{threeparttable}

\end{table*}

\begin{table*}
	\hspace{3cm}
	\caption{Same as Table~\ref{tableyields0001} for Z=0.002 
	} 
	\scalebox{1.00}{ 
	\centering                                
	\begin{tabular}{| c | c | c | c | c | c | c || c | c |}    
	\hline
	&                         &                        &                          &                           & \multicolumn{2}{c||}{Mass fraction $^{3}$He}  & &\\
	M      &                 & V$_{\rm ZAMS}$/V$_{crit}$ & V$_{\rm ZAMS}$ & life time at TO & 1DUP & 2DUP & Yield $^{3}$He  &  \\
	(M$_{\odot}$) & &                      &        (km.sec$^{-1}$)    &    (yr)                 &              &             & (M$_{\odot}$)             & \\
	\hline \hline
	0.85 & stand. & -  & - & $1.24.10^{10}$ & $1.63.10^{-03}$ & $1.33.10^{-03}$ & $6.40.10^{-04}$ & A  \\
		& thermoh & - & - & $1.24.10^{10}$& $1.62.10^{-03}$ & $1.64.10^{-04}$&$7.69.10^{-05}$ & A \\
	         & th. +rot. & 0.45 & 114 & $1.23.10^{10}$& $1.89.10^{-03}$ &$1.76.10^{-04}$ &$9.20.10^{-05}$ & A \\
          \hline
	1.0   & stand      & - &- & $6.72.10^{09}$ &$1.15.10^{-03}$ & $9.84.10^{-04}$ &  ($5.82.10^{-04}$, A) $5.66.10^{-04}$ & C \\
	         & thermoh & - & -&  $6.72.10^{09}$ & $1.15.10^{-03}$&$1.84.10^{-04}$ &($1.19.10^{-04}$, A) $1.06.10^{-04}$  & C \\
	         & th. +rot.   & 0.45 & 112 & $6.92.10^{09}$& $1.27.10^{-03}$& $2.26.10^{-04}$  & $1.32.10^{-04}$ & A \\ 
          \hline
	1.25 & thermoh & - &- & $2.88.10^{09}$ &$7.12.10^{-04}$& $2.43.10^{-04}$ & $1.71.10^{-04}$& A \\
	         & th. +rot.   & 0.45 & 115 & $2.90.10^{09}$ & $6.30.10^{-04}$&$2.45.10^{-04}$ & $1.72.10^{-04}$& A \\
          \hline
	1.5   & stand      & - & -& $1.59.10^{09}$ &$5.03.10^{-04}$&$4.55.10^{-04}$ & ($4.16.10^{-04}$, A) $4.02.10^{-04} $& C\\
	         & thermoh & - & -& $1.52.10^{09}$ &$4.90.10^{-04}$&$2.56.10^{-04}$ &   $2.18.10^{-04}$ & A \\
	         & th. +rot.   & 0.45 & 123 &$1.57.10^{09}$ &$4.59.10^{-04}$&$2.53.10^{-04}$  & $2.15.10^{-04}$& A \\
         \hline
          2.0  & stand.     & - & -& $7.14.10^{08}$& $2.95.10^{-04}$& $2.79.10^{-04}$& $3.26.10^{-04}$ & A \\
                  & thermoh & - & -& $7.14.10^{08}$ &$2.95.10^{-04}$ &$2.28.10^{-04}$ & ($2.80.10^{-04}$, A) $2.62.10^{-04}$ & C \\
                  & th. +rot.   & 0.45 & 137  & $7.21.10^{08}$& $2.61.10^{-04}$ & $2.13.10^{-04}$& $2.40.10^{-04}$ & A\\
          \hline
         2.5   & stand      & - & -& $3.91.10^{08}$ & $2.0.10^{-04}$ & $1.90.10^{-04}$& $ 2.70.10^{-04}$& A  \\
                  & thermoh & - & -& $3.91.10^{08}$ &$1.95.10^{-04}$ &$1.87.10^{-04}$ & $2.65.10^{-04}$&  A\\
                  & th. +rot.   & 0.45 & 146 & $4.03.10^{08}$&$1.66.10^{-04}$ &$1.56.10^{-04}$ & $ 2.13.10^{-04}$&  A\\
          \hline
         3.0   & stand.  & - & - &  $2.48.10^{08}$& $1.57.10^{-04}$ &$1.45.10^{-04}$ & $2.37.10^{-04}$ & A\\
         		& thermoh & - &- & $2.48.10^{08}$  &$1.57.10^{-04}$&$1.44.10^{-04}$ & $2.35.10^{-04}$ & A  \\
                  & th. +rot.   & 0.45 & 153 & $2.57.10^{08}$ &$1.26.10^{-04}$&$1.17.10^{-04}$ & $1.80.10^{-04}$& A \\       
         \hline 
         4.0 & stand.  & - & - & $1.28.10^{08}$ &$1.18.10^{-04}$& $9.79.10^{-05}$  & $1.93.10^{-04}$& A \\
         		& thermoh & - & - &  $1.28.10^{08}$ &$1.19.10^{-04}$& $1.00.10^{-4}$ & $1.99.10^{-4}$ & A \\
                  & th. +rot.  & 0.45 & 163 & $1.33.10^{08}$ &$8.80.10^{-05}$& $7.52.10^{-05}$ & $-7.05.10^{-05}$ &  B\\
         \hline
         6.0  & stand.  & - & - & $5.53.10^{07}$ &$8.17.10^{-05}$& $5.71.10^{-05}$  & -& A\\
         		& thermoh & - & - &  $5.53.10^{07}$ &$8.19.10^{-05}$& $5.72.10^{-05}$ & -  & A \\
                  & th. +rot.  & 0.45 & 170  & $5.72.10^{07}$&$6.0.10^{-05}$ & $4.40.10^{-05}$ & $-1.01.10^{-04}$  & B\\ 
        \hline
	\end{tabular} }
	\label{tableyields002}
\end{table*}

\begin{table*}
	\hspace{3cm}
	\caption{Same as Table~\ref{tableyields0001} for Z=0.004
	} 
	\scalebox{1.00}{ 
	\centering	                                
	\begin{tabular}{| c | c | c | c | c | c | c || c | c | }  
	\hline
	&                         &                        &                          &                           & \multicolumn{2}{c||}{Mass fraction $^{3}$He}  & &\\
	M      &                 & V$_{\rm ZAMS}$/V$_{crit}$ & V$_{\rm ZAMS}$ & life time at TO & 1DUP & 2DUP & Yield $^{3}$He  &  \\
	(M$_{\odot}$) & &                      &        (km.sec$^{-1}$)     &    (yr)                 &              &             & (M$_{\odot}$)             & \\
	\hline \hline
	1.0   & stand       & -& -&  $7.64.10^{09}$& $1.23.10^{-03}$ &$1.09.10^{-03}$ & $(6.39.10^{-04}$, A)  $6.29.10^{-04}$ & C  \\
	         & thermoh & -& -& $7.64.10^{09}$ &$1.23.10^{-03}$ & $2.02.10^{-04}$  & ($1.35.10^{-04}$, A)  $1.25.10^{-04}$ & C \\
	         & th. +rot.  & 0.45 & 112 & $7.69.10^{09}$&$1.40.10^{-03}$ &  $2.45.10^{-04}$  & $1.50.10^{-04}$ & A \\ 
          \hline
	1.25 & thermoh & - & -& $3.19.10^{09}$ &$7.64.10^{-04}$ & $2.72.10^{-04}$  & $1.96.10^{-04}$& A  \\
	         & th. +rot.  & 0.45& 111 & $3.49.10^{09}$&$7.27.10^{-04}$&  $2.81.10^{-04}$  & $2.04.10^{-04}$& A \\
          \hline
	1.5   & stand      & - & -& $1.66.10^{09}$&$5.25.10^{-04}$& $4.97.10^{-04}$  & $4.41.10^{-04}$& A  \\
	         & thermoh & - & -& $1.66.10^{09}$   &$5.25.10^{-04}$& $2.61.10^{-04}$  &($2.44.10^{-04}$, A) $2.26.10^{-04}$  & C \\
	         & th. +rot.   & 0.45 & 119 & $1.76.10^{09}$&$4.67.10^{-04}$ & $2.69.10^{-04}$   & $2.31.10^{-04}$ & A\\
	 \hline
          2.0  & stand.     & - &- & $7.54.10^{08}$ & $3.06.10^{-04}$& $2.96.10^{-04}$  & $3.47.10^{-04}$& A\\
                  & thermoh & - & -&  $7.54.10^{08}$ & $3.06.10^{-04}$&$2.52.10^{-04}$  & ($3.11.10^{-04}$, A)  $ 2.93.10^{-04}$& C \\
                  & th. +rot.   & 0.45 & 123 & $7.98.10^{08}$&$2.78.10^{-04}$  & $2.35.10^{-04}$  & $2.68.10^{-04}$& A\\
          \hline
         2.5   & stand       & - & -& $4.22.10^{08}$& $2.05.10^{-04}$  & $2.00.10^{-04}$  & $2.85.10^{-04}$ & A  \\
                  & thermoh & - & -& $4.22.10^{08}$ & $2.06.10^{-04}$& $2.00.10^{-04}$  & $2.85.10^{-04}$& A \\
                  & th. +rot.   & 0.45 & 141 & $4.43.10^{08}$&$1.80.10^{-04}$ & $1.76.10^{-04}$  & $2.45.10^{-04}$& A \\
          \hline
         3.0   & stand.  & - & - & $2.69.10^{08}$  &$1.57.10^{-04}$ & $1.50.10^{-04}$  & $2.46.10^{-04}$ & A \\
         		& thermoh & - & -& $2.68.10^{08}$ &$1.57.10^{-04}$& $1.49.10^{-04}$  & $2.43.10^{-04}$& A \\
                  & th. +rot.   & 0.45 & 147 & $2.80.10^{08}$ &$1.32.10^{-04}$& $1.26.10^{-04}$  & $1.98.10^{-04}$& A\\
         \hline 
         4.0 & stand.  & - & - & $1.25.10^{08}$ &$1.11.10^{-04}$& $1.03.10^{-04}$  & $2.05.10^{-04}$& A \\
         		& thermoh & - & - &  $1.25.10^{08}$ &$1.11.10^{-04}$& $1.02.10^{-04}$ & $2.04.10^{-04}$&  A\\
                  & th. +rot.  & 0.44 & 147 & $1.30.10^{08}$ &$8.81.10^{-05}$& $7.90.10^{-05}$ & $-6.13.10^{-05}$ & B \\
         \hline
         6.0  & stand.  & - & - & $5.69.10^{07}$&$7.38.10^{-05}$ & $5.98.10^{-05}$  & - & A\\
         		& thermoh & - & - &  $5.68.10^{07}$ &$7.40.10^{-05}$& $5.99.10^{-05}$ & -& A \\
                  & th. +rot.  & 0.45 & 167  & $5.88.10^{07}$ &$5.60.10^{-05}$& $4.59.10^{-05}$ & $-9.77.10^{-05}$ & B \\
         \hline
	\end{tabular} }
	\label{tableyields004}

\end{table*}
	
\begin{table*}
	\hspace{3cm}
	\caption{Same as Table~\ref{tableyields0001} for Z=0.014
	} 
	\scalebox{1.00}{ 
	\centering
	                              
	\begin{tabular}{| c | c | c | c | c | c | c || c | c |}      
	\hline
	&                         &                        &                          &                           & \multicolumn{2}{c||}{Mass fraction $^{3}$He}  & &\\
	M      &                 & V$_{\rm ZAMS}$/V$_{crit}$ & V$_{\rm ZAMS}$ & life time at TO & 1DUP & 2DUP & Yield $^{3}$He  &  \\
	(M$_{\odot}$) & &                      &        (km.sec$^{-1}$)    &    (yr)                 &              &             & (M$_{\odot}$)             & \\
	\hline \hline
	1.0   & stand      & - & - & $1.15.10^{10}$ & $1.45.10^{-03}$&$1.28.10^{-03}$ & $7.73.10^{-04}$ & A \\
	         & thermoh & - & -  &$1.15.10^{10}$ &$1.44.10^{-03}$& $2.79.10^{-04}$ &($1.97.10^{-04}$, A) $1.92.10^{-04}$  & C  \\
          \hline
	1.25 & thermoh & - & - & $4.70.10^{09}$ &$8.74.10^{-04}$& $3.33.10^{-04}$ &  ($2.61.10^{-04}$, A) $2.54.10^{-04}$ & C\\
	         & th. +rot.  & 0.45 & 110 & $5.04.10^{09}$ &$9.33.10^{-04}$& $2.93.10^{-04}$ & ($2.43.10^{-04}$, A) $2.39.10^{-04}$  & C\\
          \hline
	1.5   & stand      & - & - & $2.29.10^{09}$ &$6.04.10^{-04}$ &$5.85.10^{-04}$ & ($5.21.10^{-04}$, A) $5.20.10^{-04}$  & C \\
	         & thermoh & - & - & $2.29.10^{09}$ & $6.04.10^{-04}$&$3.77.10^{-04}$ & ($3.53.10^{-04}$, A) $3.43.10^{-04}$  & C\\
	         & th. +rot.  & 0.45 & 110 & $2.56.10^{09}$ &$5.68.10^{-04}$& $2.95.10^{-04}$ & ($2.70.10^{-04}$, A)  $2.64.10^{-04}$  & C\\ 
	 \hline
          2.0  & stand.     & - & - & $1.01.10^{09}$ &$3.40.10^{-04}$& $3.33.10^{-04}$ & ($3.93.10^{-04}$, A)  $ 3.92.10^{-04}$ & C \\
                  & thermoh & - & - & $9.61.10^{08}$ &$3.40.10^{-04}$& $3.06.10^{-04}$ & $3.62.10^{-04}$ & A \\
                  & th. +rot.  & 0.40 & 110 & $1.08.10^{09}$ &$3.19.10^{-04}$& $2.73.10^{-04}$ &($3.18.10^{-04}$, A)  $3.12.10^{-04}$  & C\\
                  & th. +rot.  & 0.91 & 250 & $1.08.10^{09}$ & $2.96.10^{-04}$ & $2.6.10^{-04}$ & $2.97.10^{-04}$& A\\
          \hline
         2.5   & stand       & - & - & $5.22.10^{08}$ & $2.19.10^{-04}$ &$2.18.10^{-04}$ & $3.12.10^{-04}$ & A \\
                  & thermoh & - & - & $5.22.10^{08}$ &$2.19.10^{-04}$& $2.19.10^{-04}$ & $3.14.10^{-04}$&  A\\
                  & th. +rot. & 0.45 & 130 & $5.76.10^{08}$ &$2.03.10^{-04}$& $2.02.10^{-04}$ & ($2.86.10^{-04}$, A) $2.80.10^{-04}$ & C\\
          \hline
         3.0   & stand. & - & - & $3.37.10^{08}$ & $1.63.10^{-04}$ &$1.61.10^{-04}$ & $ 2.65.10^{-04}$  & A \\
         		& thermoh & - & - & $3.21.10^{08}$ &$1.60.10^{-04}$ & $1.58.10^{-04}$ & $2.59.10^{-04}$ & A \\
                  & th. +rot.  & 0.45 & 136 & $3.52.10^{08}$&$1.48.10^{-04}$  & $1.46.10^{-04}$ & ($2.35.10^{-04}$, A)  $2.30.10^{-04}$  & C \\
         \hline 
         4.0 & stand.  & - & - & $1.61.10^{08}$ &$1.07.10^{-04}$ & $1.04.10^{-04}$  & $2.04.10^{-04}$  & A \\
         		& thermoh & - & - & $1.61.10^{08}$ &$1.07.10^{-04}$ & $1.03.10^{-04}$ & $2.04.10^{-04}$ & A \\
                  & th. +rot.  & 0.45 & 144 & $1.67.10^{08}$ &$9.41.10^{-05}$ & $9.18.10^{-05}$ &   $1.72.10^{-04}$& A \\
                   & th. +rot.  & 0.93 & 300 &  $1.68.10^{08}$  & $8.41.10^{-05}$ & $8.23.10^{-05}$ &   $1.46.10^{-04}$& A \\
         \hline
         6.0  & stand.  & - & - & $6.16.10^{07}$ &$6.80.10^{-05}$ & $5.92.10^{-05}$  & $-1.05.10^{-04}$  & B \\
         		& thermoh & - & - &  $6.16.10^{07}$ &$6.75.10^{-05}$ & $5.89.10^{-05}$ & -  &  A \\
                  & th. +rot.  & 0.45 & 156 & $7.24.10^{07}$&$5.78.10^{-05}$  & $5.04.10^{-5}$ &  $-5.37.10^{-05}$& B\\
          \hline
	\end{tabular}}
	\label{tableyields014}

\end{table*}

%============================================================= Conclusions =======================================================

\begin{acknowledgements}
 
We thank J.-P.Zahn, C.Chiappini, D.Romano, M.Tosi, T.Bania, R.T.Rood and D.Balser for fruitful discussions on the $^3$He problem over the years.
We acknowledge financial support from the Swiss National Science Foundation (FNS), from ESF-Euro Genesis,  and the french Programme National de Physique Stellaire (PNPS) of CNRS/INSU.
\end{acknowledgements}

\bibliographystyle{aa}
\bibliography{Reference}

\end{document}